\documentclass[aps,prl, twocolumn,amsmath,amssymb,longbibliography,superscriptaddress]{revtex4-2}
\usepackage{amsmath,amssymb,amsfonts,bm}
\usepackage{graphicx}
\usepackage{xcolor}
\usepackage{bbold}
\usepackage{graphicx}
\usepackage{dcolumn}
\usepackage{epstopdf}
\usepackage{epsfig}
\usepackage[caption=false]{subfig}
\usepackage{url}
\usepackage{hyperref}
\usepackage{verbatim}
\usepackage[export]{adjustbox}
\usepackage{scalerel}
\usepackage{braket}

\usepackage{ulem}

\usepackage{amsthm}
\usepackage{enumitem}

\usepackage{multirow}
\usepackage{booktabs}
\usepackage{pifont}

\usepackage{colortbl}

\newcommand{\Nb}{N_{\mathrm{o}}}

\usepackage{float}
\newcommand{\Tr}{\operatorname{Tr}}

\newcommand{\Trt}{\operatorname{Tr}_{\mathcal{H}}}
\newcommand{\Trs}{\operatorname{Tr}_{\tilde{\mathcal{H}}}}

\newcommand{\dsff}{\mathcal{K}}
\newcommand{\dsffc}{\dsff_{\mathrm{c}}}
\newcommand{\dsffgin}{\dsff_{\mathrm{Gin}}}
\newcommand{\dsffcgin}{\dsff_{\mathrm{c}, \mathrm{Gin}}}

\newcommand{\KGinUE}{K_{\mathrm{Gin}}}
\newcommand{\kappaginue}{\kappa_{\mathrm{Gin}}}

\newcommand{\kappaginueB}{\kappa_{\mathrm{F}-\mathrm{Gin}}}

\newcommand{\kappaginueC}{\kappa_{\mathrm{TI}-\mathrm{Gin}}}

\newcommand{\kk}{k}

\newcommand{\be}{\begin{equation}}
\newcommand{\ee}{\end{equation}}
\newcommand{\ba}{\begin{aligned}}
\newcommand{\ea}{\end{aligned}}

\newcommand{\bmult}{\begin{multline}}
\newcommand{\emult}{\end{multline}}

\newcommand{\blue}[1]{{\color{blue} #1}}

\newcommand{\Ksff}{K}

\newcommand{\kd}{K_\mathrm{TIF}}

\newcommand{\kappac}{\kappa_\mathrm{TI}}

\newcommand{\hJ}{\hat{J}}

\newcommand{\hh}{\hat{h}}
\newcommand{\tomega}{\tilde{\omega}}

\newcommand{\bs}{\mathbf{s}}
\newcommand{\bp}{\mathbf{p}}

\def\tth{t_{\rm Th}}
\def\Lth{L_{\rm Th}}

\def\ts{t^*}

\newcommand{\Ls}{L^*}
\newcommand{\Lsb}{L^*_{\mathrm{F}}}
\newcommand{\Lsc}{L^*_{\mathrm{TI}}}

\newcommand{\kfgin}{K_{\mathrm{F-Gin}}}
\newcommand{\ktifgin}{K_{\mathrm{TIF-Gin}}}
\newcommand{\kappafgin}{\kappa_{\mathrm{F-Gin}}}

\def\bv{{\bf b}}

\def\varr{{v}}

\def\VV{V}

\newcommand{\Jc}{J_{\mathrm{c}}}

\graphicspath{{Figures/}}

\newcommand{\average}[1]{\left\langle #1 \right\rangle}

\def\VG{V_{\rm G}}

\begin{document}

\title{Many-body quantum chaos and emergence of  Ginibre ensemble} 
\newcommand{\titleinfo}{Many-body quantum chaos and emergence of  Ginibre ensemble}

\author{Saumya  Shivam}
\affiliation{Department of Physics, Princeton University, Princeton, New Jersey 08544, USA}

\author{Andrea De Luca}
\affiliation{Laboratoire de Physique Th\'eorique et Mod\'elisation, CY Cergy Paris Universit\'e, \\
\hphantom{$^\dag$}~CNRS, F-95302 Cergy-Pontoise, France}    

\author{David A. Huse}
\affiliation{Department of Physics, Princeton University, Princeton, New Jersey 08544, USA}

\author{Amos Chan}
\affiliation{Department of Physics, Lancaster University, Lancaster LA1 4YB, United Kingdom}
\affiliation{Princeton Center for Theoretical Science, Princeton University, Princeton NJ 08544, USA}

\date{\today}

\begin{abstract}

We show that non-Hermitian Ginibre random matrix  behaviors emerge in spatially-extended many-body quantum chaotic systems in the space direction, just as Hermitian random matrix behaviors emerge in chaotic systems in the time direction.
Starting with translational invariant models, which can be associated with dual transfer matrices with complex-valued spectra, we show that the linear ramp of the spectral form factor necessitates that the dual spectra have non-trivial correlations, which in fact fall under the universality class of the Ginibre ensemble, demonstrated by computing the level spacing distribution and the dissipative spectral form factor.
As a result of this connection, the exact spectral form factor for the Ginibre ensemble can be used to universally describe the spectral form factor for translational invariant many-body quantum chaotic systems in the scaling limit where $t$ and $L$ are large, while the ratio between $L$ and $\Lth$, the many-body Thouless length is fixed. 
With appropriate variations of Ginibre models, we analytically demonstrate that our claim generalizes to models without translational invariance as well.
%
%
The emergence of the Ginibre ensemble is a genuine consequence of the strongly interacting and spatially extended nature of the quantum chaotic systems we consider, unlike the traditional emergence of Hermitian random matrix ensembles.

\end{abstract}

\maketitle

\begin{figure}[t!]
\centering
\includegraphics[width=0.44\textwidth,trim=0.22cm 0.1cm 0.22cm 0cm]{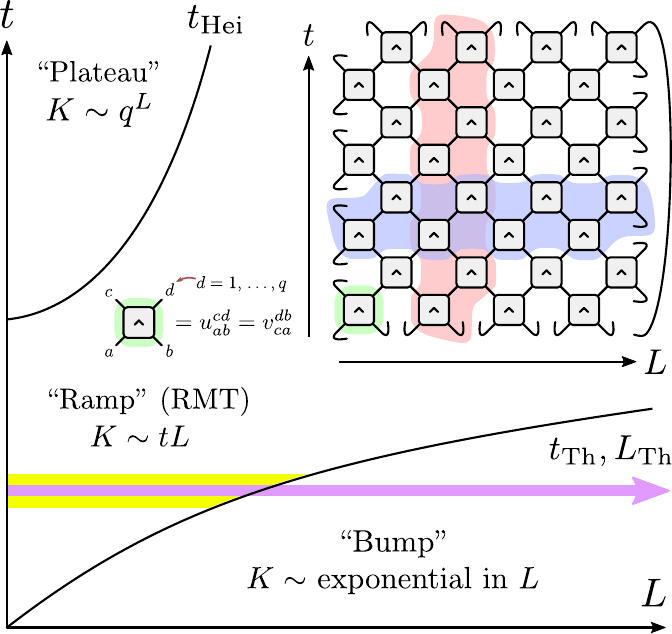}
\caption{Regime diagram of spectral form factor $K(t,L)$ (Eq \ref{eq:sff_dual}) for many body quantum chaotic systems with translational invariance in space and time, with `Bump', random matrix `Ramp' (RMT) and `Plateau' regimes. 
For fixed $t$ and increasing $L$ (purple), the SFF exhibits an initial linear ramp behavior (yellow) which necessarily requires non-trivial spectral statistics of the dual spectra.
Inset:
Diagrammatical representation of equality of the spectral form factor computed using the dual transfer matrix (Eq \ref{eq:sff_dual}) , with unitary 2-gate (green), Floquet operator $W(L)$ (blue), dual transfer matrix $V(t)$ (red).
\label{Fig:sff_reg}}
\end{figure}

\textit{Introduction.}--
%
%
The discovery of the connection between quantum chaos and random matrix theory (RMT) is of great importance in theoretical physics
because RMT provides an approach that eliminates dependence on the microscopic details 
and captures the universal characteristics of an ensemble of statistically similar chaotic systems, constrained only by symmetries~\cite{Mehta, Haake}.
%
Historically, the  spectral correlation of the Gaussian ensembles was discovered in chaotic
mesoscopic systems for sufficiently small energy scales or equivalently, sufficiently late time scales~\cite{bohigas1984characterization, AltshulerShklovskii}.
%
%
%
Recently, with the developments in random unitary circuits~\cite{Nahum2017, PhysRevB.99.174205, Li_2018, Skinner_2019, Li_2019, chan2019, Gullans_2020, altman2019, Jian_2020, Zabalo_2020,Nahum2017a, vonKeyserlingk2017, vonKeyserlingk2017a, Huse2017}, particularly in time periodic or Floquet circuits, 
 analytic calculations of random matrix behaviour in spectral correlations of spatially-extended \textit{many-body} quantum chaotic systems have been achieved~\cite{cdc1, cdc2, Prosen, bertini2018exact,chan2021manybody, cdc3, friedman2019, cdclyap, moudgalya2021}. %
%
%
%
%
%
While Floquet circuits have given access to the study of non-trivial spectral properties in extended many-body systems --- like the onset of RMT behaviour~\cite{ cdc2, friedman2019, moudgalya2021,Gharibyan_2018, Saad2019semiclassical}, spectral Lyapunov exponents~\cite{cdclyap}, and novel scaling forms and limits~\cite{friedman2019, chan2021manybody} --- 
%
%
%
translational-invariant (TI) circuits give rise, via the so-called space-time duality, to non-Hermitian dual transfer matrix (Fig.~\ref{Fig:sff_reg} red) 
with complex eigenvalues, the dual spectrum. 
%
%
The study of many-body quantum system using space-time duality began in the study of the kicked Ising model at the self-dual point \cite{guhr2016kim, bertini2018exact, bertini2021random, waltner2019dual, guhr2020kim} and concurrently  in the transfer matrix approach in Floquet circuits \cite{cdc2, friedman2019, cdclyap}.
%
Subsequently, numerous works have investigated the non-unitary ``dynamics'' in the space direction~\cite{lerose2020influence,  ippoliti_2021_dual, ippoliti2021fractal, grover2021entanglement, zhou2021spacetime}.
The objective of this paper is to provide evidence of the emergence of non-Hermitian Ginibre (GinUE) RMT-behaviour~\footnote{The non-Hermitian GinUE RMT universality class is also known as non-Hermitian class A~\cite{hamazaki_universality_2020}.  The label ``U'' in GinUE indicates that the distribution is invariant under unitary conjugation, although the Ginibre matrices themselves are non-unitary.}  
in many-body quantum chaotic (MBQC) systems in the thermodynamic and scaling limit, in contrast to the emergence of standard Gaussian Hermitian RMT ensembles in late time,
as illustrated below. 
{
\begin{figure}[h]
\centering
\includegraphics[width=0.467\textwidth]{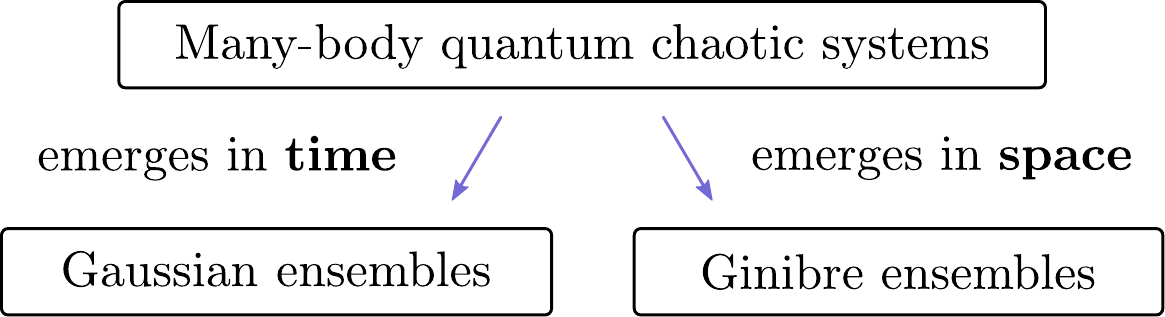}
\end{figure}
}

%
%
%
%

%
%
%

%




\begin{figure*}[ht!]
\centering
\includegraphics[width=0.98\textwidth,trim=0cm 0.30cm 0cm 0cm]{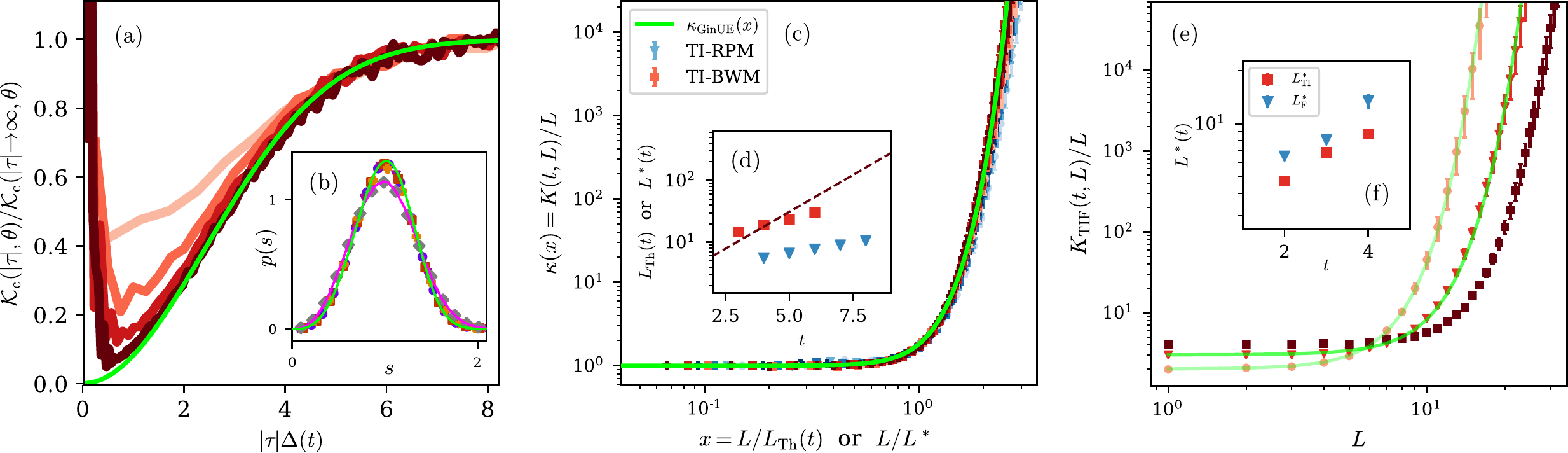}
\caption{
%
%
%
%
Universal correlations for representative 
many body random quantum circuits~\cite{supplementary}, showing approach to corresponding quantities computed for the Ginibre ensemble (green).
(a): Dissipative spectral formm factor (\ref{eq:dsff_def_a}) of the dual spectra of the brick wall model, for $t=3,4,5,6$ from light to dark red.
(b): Nearest neighbour spacing distribution of the dual spectra of the brick wall model (on site dimension $q=2,t=6$, purple), random phase model ($q=3,t=8$, burgundy), and zero momentum sectors of
translational invariant Floquet brick wall model ($q=2,t=7$, red) and random phase model ($q=3,t=10$, gold). Kicked Ising model away from the self dual point at $J=0.75J_c$ (grey) shows the distribution corresponding to the symmetric Ginibre ensemble (pink curve obtained from $N=2187$) due to time reversal symmetry~\cite{supplementary}.
(c): Scaling collapse of the spectral form factor $ \kappac(x)$ for two models and  $ \kappaginue(x)$, for the Ginibre ensemble (\ref{eq:conjTI}), against $x = L/\Lth$ or  $L/ \Ls$ with excellent agreement, where $\Lth$ is Thouless length, and  $\Ls$ is the inverse mean level spacing for Ginibre ensemble. 
(d): $L/\Lth$ (dots) and $\Ls$ (dashed line) against time $t$, used for the collapse in the main panel. For Ginibre, we define an effective time via $N:= q^t$.
%
(e): 
Scaled spectral form factor $\kd(t,L)/L$  for translational invariant Floquet brick wall model ($q=3, t=2,3,4$, red) and the numerical fit of $\ktifgin(t,L)/L$ (green)  against $L$ with darker colors for larger $t$. We fit $\ktifgin(t,L)$ to $\kd(t,L)$ by tuning $\Lsb$  and $\Lsc$ in Eq.~\eqref{eq:tif_k}, which are plotted against time $t$ as blue and red respectively in (f).}
%
%

 \label{fig:three_panels}
\end{figure*}


\paragraph{Heuristics.--}
One of the simplest non-trivial and analytically-tractable quantities to diagnose chaos
is the \textit{spectral form factor} (SFF), defined as  \cite{Haake, cdc1, cdc2, friedman2019, cdclyap,  moudgalya2021, chan2021manybody, bertini2018exact, bertini2021random, flack2020statistics, Prosen,  Cotler_2017, complexity2017, Saad2019semiclassical, Gharibyan_2018, ALTLAND2018, garratt2020dw, garratt2020MBL, li2021spectral, ZollerSFF2020, ZollerSFF2021, prakash2021sffmbl, winer2020exp, santos2022boson, Liao2020, liao2022emergence, Cornelius2022}
\begin{equation}\label{eq:sff_dual}
\Ksff(t,L) 
=
 \average{ \left| \Trt\left[ \mathcal{W}(t,L)\right] \right|^2}
 =
  \average{ \left| 
  \Trs\left[ \mathcal{V}(t,L)\right]\right|^2}
\end{equation}
where $\mathcal{W}(t,L) = \prod_{t'=1}^{t} W(t', L)$  is a time evolution operator acting on Hilbert space $\mathcal{H}$, and $\mathcal{V}(t, L) = \prod_{j=1}^L V(t, j)$ is the corresponding dual operator (Fig.~\ref{Fig:sff_reg} red) performing  ``evolution'' in space on dual Hilbert space $\tilde{\mathcal{H}}$. $t$ and $L$ denote the numbers of repeated actions of $W$ and $V$, and can be treated as effective time and system size respectively \footnote{\label{note1} The dimensions of the Hilbert space that the operator acts on in general depends on the model. For the brick-wall model in Fig. \ref{Fig:sff_reg} (where $t=4,L=4$), it is $q^{2t}$ for the time transfer matrix and $q^{2L}$ for the space transfer matrix, $q$ being the on-site dimension. On the other hand for the random phase model and the kicked Ising model, it is $q^t$ for the time transfer matrix and $q^L$ for the space transfer matrix. } . 
For Floquet systems, one has $W(t',L) = W(L)$, while $V(t,j) = V(t)$ for TI systems with transfer matrix $V(t)$. We can generally diagonalise $V(t)$ with the eigenvalues $\{z_j \} \equiv\{\rho_j e^{\imath \phi_j}\}$ with $\rho_j, \phi_j \in \mathbb{R}$. \begin{equation}
\label{eq:SFFdual}
K(t,L) =
  \average{ 
 \sum_{i} \rho_i^{2L}
 +
\sum_{i\neq j} 
\left[ 
e^{i \left(\phi_i - \phi_j \right)}  \rho_i \rho_j  
\right]^L
}
\end{equation}
We are denoting as $\average{\ldots}$  the ensemble average over statistically similar systems.
In the absence of extra symmetries, RMT predicts $K(t, L) \sim t L$ for TI Floquet systems. This can be understood as the spectrum of $W(L)$ splits in $L$ momentum sectors which emerges because of TI. If correlations between sectors vanish, the spectral form factor results from the sum of the usual linear-in-$t$ behavior within each momentum sector~\cite{chan2021manybody}. 
For many-body systems, this RMT behavior emerges whenever $t > \tth(L)$ or equivalently $L < \Lth(t)$, where $\tth(t)$, $\Lth(L)$ are respectively the many-body SFF Thouless time and length, related by $\Lth(\tth(L)) = L$. The Thouless time is a system-dependent quantity which characterises the time scale for the onset of chaos in the two-point level correlation and in general is expected to grow with system size $L$~\cite{chan2021manybody} (with the relevant exception of the dual--unitary circuits~\cite{Akila_2016, bertini2018exact, bertini2021random, flack2020statistics, ippoliti2021fractal, lamacraft2019, influencematrix}).
It is insightful to re-interpret these considerations in terms of the spectrum of $V(t)$. From Eq.~\eqref{eq:SFFdual}, we see that if phase correlations could be neglected, $K(t,L) \gtrsim e^{\lambda(t) L}$, with $\lambda(t) = \max_i \ln \rho_i$ for $L \gg \Lth(t)$. We label this regime as the ``Exponential bump'' region in Fig.~\ref{Fig:sff_reg}. Thus, the existence of the ``Ramp'' regime, characteristic of RMT, for $L \lesssim \Lth(t)$ implies
that the off-diagonal term in \eqref{eq:SFFdual} \textit{necessarily} display non-trivial correlation, such that the exponential behaviour of the diagonal term in \eqref{eq:SFFdual} could be compensated. 
We emphasize that this heuristic argument applies to generic translational invariant MBQC systems. The characterisation of the spectral statistics of  $V(t)$ will be the main objective of this letter. As we show below,
such dual spectral statistics falls under the universality class of Ginibre ensemble, which can be seen as the most generic rotation invariant Gaussian ensemble, once all relevant symmetries have been taken into account (e.g. space-time translational invariance). 

\paragraph{Models.}--
We consider three one-dimensional random unitary circuits as models of MBQC, namely the brick-wall model, the random phase model, and the kicked Ising model. 
All three models can be written  as the operator $\mathcal{W}(t,L)= \prod_{t'=1}^t W(t',L) = \prod_{r=1}^L V(t, r) =\mathcal{V}(t,L)$ where $W(t',L)$ and $V(t,r)$ refer respectively to the time and space transfer matrix shown in blue and red in Fig.~\ref{Fig:sff_reg}, acting on the Hilbert space with dimensions $q^L$ and $q^t$ respectively, with $q$ being the on-site dimension \footnotemark[\value{footnote}]. The circuit is composed of unitary two-gates $u(t',r)$ and one can define the space-time dual of $u$ via $u^{cd}_{ab}(t',r) =v^{db}_{ca}(t',r)$. The precise definitions of the gates $u(t',r)$  
are given in the supplementary material \cite{supplementary}, and are not crucial for our discussion as long as the models are chaotic and have no conserved quantities. 
We define four setups resulting from the combination of translational invariance in space and time: 
(a) Temporally and spatially random unitary circuits, where all $u$-s are drawn independently. In this case, spectral correlations are trivial in both space and time directions{, with $K(t,L)\sim 1$ for all $t$, $L$ \cite{chan2021manybody}};
%
(b) Temporally periodic, i.e. Floquet, and spatially random (Floquet) circuits, where $u(t',r)= u(t'',r)$ for all  $t'$, $t''$ and $r$; 
(c) Temporally random and spatially TI 
random circuits, where $u(t',r)= u(t',r')$ for all $t'$, $r$ and $r'$; 
and (d) Floquet and spatially TI 
(TIF) circuits, where $u(t',r)= u(t'',r')$ for all $t', t'', r$ and $r'$.
%
%
%
%

\textit{Dual spectral statistics.}-- 
We start by focusing on TI {(temporal random) models (case c)}, where the transfer matrix $V(t)$ has a well-defined spectrum and exhibits no additional symmetries since the model is temporarily disordered.  As the spectrum is complex, in order to analyse its correlations, we resort to a) level spacing distribution and b) a natural generalization of SFF, known as the \textit{dissipative spectral form factor} ~\cite{li2021spectral}.
The SFF of a generic complex spectrum is exponentially growing or decaying due to the imaginary parts of the complex eigenvalues. %
To circumvent this problem, dissipative SFF instead treats the complex spectrum as a set of points in the plane and assess the distribution of their euclidean distances. Indeed, for a non-Hermitian operator with spectrum $\{z_n = x_n + iy_n: x_n, y_n \in \mathbb{R} \}$,
the connected part is defined as 
\begin{equation} \label{eq:dsff_def_a}
\dsff_\mathrm{c}(t,s):= 
\left\langle
\left|
\sum_{n} e^{ix_n t + iy_n s}
\right|^2
\right\rangle
-
\left|
\left\langle
\sum_{n} e^{ix_n t + iy_n s}
\right\rangle
\right|^2
\;,
\end{equation}
where $t$ and $s$ are two generalized time variables. We organise them into the complex time $\tau \equiv  t+ is \equiv |\tau | \, e^{\imath \theta}$, and will abusively use the polar coordinate $(|\tau|,\theta)$ to parameterise the arguments of $\mathcal{K}_c$.
As a yardstick for the generic behavior of $\mathcal{K}_c$, we consider the GinUE, sampled by taking $N$-by-$N$ random matrices with independent complex Gaussian matrix element with variance $\sigma^2 = \varr /N$. In other words, the probability density for a matrix $M$ is $\propto \exp [ - N/(2v) \Tr MM^\dag]$, and is thus rotational invariant.
Therefore, the GinUE is expected to capture the spectral correlations of sufficiently generic, or ``chaotic'', complex non-Hermitian matrices, in a similar fashion to how the Gaussian and Circular unitary ensemble are the universality class for unitary and hermitian matrices respectively~\cite{li2021spectral, upcomingli}.
The dissipative SFF can be computed explicitly for GinUE~\cite{li2021spectral}. Keeping the leading contribution in $N$, $\mathcal{K}_c$ simplifies to
\be   
\dsffcgin (|\tau| , \theta ) = \frac{N}{\varr} 
\left(
1 - e^{ -\frac{ \varr |\tau|^2}{ 4N} } \right) \,.
\label{eq:ginDSFF}
\ee
which is rotational symmetric and shows a \textit{(dip-)ramp-plateau} behaviour~\footnote{At early time $|\tau |  \ll \Delta^{-1}$, 
DSFF dips from $K(0,\theta)= N^2$ with a form described by the non-universal disconnected DSFF, 
$\left| 
\left\langle \sum_{n} e^{i x_n t + i y_n s} 
\right\rangle 
\right|^2$, discussed in \cite{li2021spectral}, but excluded in Eq. \blue{\eqref{eq:ginDSFF}} for brevity},
analogous to the SFF for closed quantum systems: 
At $ |\tau| \lesssim \Delta^{-1} \sim \sqrt{N}$,
it increases \textit{quadratically} $ \simeq |\tau|^2/4$ in large $N$
%
until it plateaus at $N$ at a time comparable to the inverse of the mean level spacing $\Delta$ in the complex plane.   
Remarkably, the quadratic ramp of dissipative SFF for GinUE is drastically different from the corresponding behaviour for Gaussian unitary ensembles, which is \textit{linear} in time. 
The quadratic ramp is sensitive to the variation of density of states across the complex plane, and thus unfolding is required to uncover the true long-range dual spectral correlations~\cite{supplementary}. %
In Fig.~\ref{fig:three_panels}a, we show for TI random phase model, as a representative example, a good collapse of $\dsffc(|\tau|, \theta) /\dsffc(|\tau|\to \infty, \theta)$ against $|\tau|\Delta$, approaching GinUE behaviour~\eqref{eq:ginDSFF} as the dual system size $t$ increases, with a similar approach for other models~\cite{supplementary}, demonstrating universality. \\

To provide further evidence of emergence of GinUE,
we probe the spectral correlation at the scale of mean level spacing in the complex plane using the nearest-neighbour spacing distribution in Fig.~\ref{fig:three_panels}b, and complex spacing ratio~\cite{sa_complex_2020} in \cite{supplementary}, for the three different TI models. 
We find signatures of level repulsion consistent with the corresponding RMT universality classes (including the ones with time reversal symmetry~\cite{supplementary}), and with the dissipative SFF results around the $\Delta^{-1}$ region.  
%
%
%
%
%
%
%
%
%
%
%
%
\paragraph{SFF of GinUE.--}
With the insight that dual-spectral correlation falls under the universality class of GinUE, it is natural to ask whether this information can be used to understand the behavior of the SFF. 
As before, we start by focusing on TI systems, where, in the absence of extra symmetries, the correlation of the dual spectrum are captured by the standard GinUE, whose joint probability distribution function of eigenvalues $\{z_j \}$ is known exactly~\cite{ginibre_1965}. 
We model the SFF in \eqref{eq:sff_dual}, by replacing the transfer matrix $V(t)$ with $\VG$ drawn from the GinUE of size $N$, and obtain~\cite{supplementary}
\be\label{eq:kginue}
\begin{aligned}
\KGinUE(N,L) := & \average{\left|\Tr\left[\VG^L\right] \right|^2} 
\\
= & N^2 \delta_{L,0}+
\frac{ \varr^L \left((L+N)!-\frac{N! (N-1)!}{ (N-L-1)!}\right)}{N^L (L+1) (N-1)!} \;.
\end{aligned}
\ee
In matching the predictions of \eqref{eq:kginue} with many-body models, we 
encode $t$ dependence in the matrix size $N$, whose functional form will be specified later. In the limit of large $N$ at fixed large $L$, $\KGinUE(N,L) = v^L L (1 + O(L^4/N^2))$. 
The crossover scale $\Lsc = \sqrt{N}$ is related to the inverse of mean level spacing $\Delta$ in the complex plane.
This suggests a scaling limit where $L$ and $N$ are sent to infinity with $x=L/\Lsc$ fixed and one has
\begin{equation} \label{eq:kc_scaling_dgt1}
    \kappaginue(x) \equiv 
      \lim_{\substack{L,N \to \infty\\ x= L/\Lsc }} 
      \frac{
      \KGinUE
      }{
      \varr^L  L
      }
      = \frac{2 \sinh \left(\frac{x^2}{2}\right)}{x^2}
  \;. 
\end{equation}
In fact, the above scaling form of GinUE
shares similarities with the scaling forms proposed for {TI (temporal random)
 systems} in \cite{chan2021manybody}, given by
\be
\begin{aligned} 
   \kappa_{\mathrm{TI}-\mathrm{MBQC}} (x)=  & \lim_{\substack{L,t \to \infty\\ x =L/L_{\rm Th}(t)}}
   \frac{
   K_{\mathrm{TI}-\mathrm{MBQC}}
     }{
     L
     }
      \;, \label{eq:sfTI}
\end{aligned}
\ee
for TI systems, where instead of $\Lsc$ in GinUE, the system-dependent many-body Thouless length $\Lth(t)$ is used to define the scaling limit. 
Now, given that (i) the spectral correlation dual spectra of many body chaotic systems falls under the GinUE universality class; 
(ii) a linear ramp in $L$ naturally emerges from \eqref{eq:kginue} for $L \ll \Lsc$, coinciding with the appearance of the linear-ramp in SFF of chaotic systems;
we conjecture that the scaling form of GinUE describes the scaling form of TI chaotic systems once $\Lsc$ and $\Lth(t)$ are identified, i.e.
\be \label{eq:conjTI}
\kappa_{\mathrm{TI}-\mathrm{MBQC}} (x)  =
\kappaginue(x) \equiv
\kappaginueC(x) \;,
\ee
To test this claim, we simulate both sides of \eqref{eq:conjTI}, for TI brick wall model, random phase model, and GinUE in Fig.~\ref{fig:three_panels}c and find an excellent collapse. 
We note that the scaling limit in \eqref{eq:sfTI} differs from the infinite-$q$ result obtained for the random phase model in \cite{chan2021manybody} which disagreed with the finite-$q$ numerical simulations.
The universality of $\kappa(x)$ implies that the microscopic details are only reflected in the function $\Lth(t)$, and not in the scaled function $\kappa(x)$, as observed in \cite{chan2021manybody}. 
Also, the validity of Eq.~\eqref{eq:conjTI} indicates that the effective size $N$ of the equivalent GinUE matrix shall not be fixed from the dual Hilbert space dimension ($= q^{2t}$), but rather from the emerging Thouless length, i.e. $N = L^{*2} \sim \Lth(t)^2 \ll q^{2t}$ (Fig.~\ref{fig:three_panels}d).

\textit{Beyond translational invariance.}-- 
We now extend the previous considerations to Floquet systems. We first consider with spatial randomness (case b) and then we incorporate TI (case d), and demonstrate the emergence of GinUE--like behaviour with and without TI. 
To incorporate time periodicity, we first 
observe that the transfer matrix becomes invariant under time translations, and thus its spectrum can be split in \textit{time momentum} or frequency sectors. In the inset of Fig.~\ref{fig:three_panels}b and~\cite{supplementary}, we respectively compute the dissipative SFF and spacing distribution for the dual spectrum in each sector, and confirm the emergence of Ginibre statistics.
Time translation implies $V(t, r) = T V(t, r)  T^{-1}$, where $T$ shifts the dual system
over one period. For simplicity, we assume invariance under one site translation, with
$T \ket{\mathbf{s} = s_1s_2\ldots s_t} = \ket{ s_2s_3\ldots s_ts_1}$, generalization to longer unit cells being straightforward.
For each configuration $\mathbf{s}$, we define its associated \textit{period} as the minimal $\tau = 1,\ldots, t$ such that $T^\tau \ket{\mathbf{s}} = \ket{\mathbf{s}}$.
To formulate the statistical properties of the ensemble, we restrict the Hilbert space to the set of  computational basis $\{ \ket{\mathbf{s}}\}$ translational invariant with only period $t$.
Indeed, the fraction of configurations with maximal period goes to $1$ for large $t$ (and/or obviously for large $q$). Using $\Nb$ to denote the number of distinct orbits under the translation operation, we formally have a dimension for the restricted dual Hilbert space $\operatorname{dim}(\tilde{\mathcal{H}}) = t \Nb$.
Then, we model the transfer matrix $V(t,r)$ by a random matrix $\VG$ with complex Gaussian entries and covariance
\be \label{eq:gin_corr}
\left\langle [\VG]_{\bs \bs'} [\VG]^
*_{\bp \bp'} \right\rangle 
=
\frac{1}{\Nb}
\sum_{\tau, \tau'}
\delta_{ \mathbf{s} T^\tau(\mathbf{p}) } \delta_{\mathbf{s}' T^{\tau'}(\mathbf{p}') }
J(\tau - \tau')\;,
\ee 
where $J(\tau - \tau')$ controls the correlation between matrix elements. 
As pointed out in \cite{Berry} via a semiclassical expansion, the emergence of SFF linear ramp using RMT $K(t) = t$ in single-particle chaotic Floquet systems can be associated to the pairing between two periodic orbits,
 which can happen in $t$ possible ways ($t$ being the discrete length of the orbit here). 
In extended chaotic systems, the factor of $t$ corresponds to the possible values of $\tau = 1,\ldots, t$ for \textit{local} pairing of orbits~\cite{cdc2, garratt2020dw}. 
%
The interaction between neighbouring local degrees of freedom forces similar pairings between local orbits, quantified here by the function $J(\tau - \tau')$. 
A simple calculation gives
$K_{\rm F-Gin}(t,L) =  \left\langle 
\left| \Tr[\VG(t,L)] \right|^2 
\right\rangle = \sum_{\{\tau\}} \prod_{i=1}^L J(\tau_i - \tau_{i+1})=\sum_\omega [\hJ(\omega)]^L$ \cite{supplementary}, with $\hJ(\omega)$ the Fourier transform of $J(\tau)$. 
We thus see that the SFF behavior in the scaling limit depends on $\hJ(\omega)$. 
For simplicity, we suppose $J(\tau - \tau') = \delta_{\tau , \tau'} + f(t) h(\tau - \tau')$, where $f(t)$ decays to zero on the scale of the Thouless time, and the function $h(\tau - \tau')$ controls the correlation between neighbouring pairings. 
Within this formulation, the scaling limit depend on the details of the Fourier transform $\hat h(\omega)$. However, the exact calculation in the random phase model at infinite $q$~\cite{cdc2, chan2021manybody, supplementary} leads to $h(\tau - \tau') = 1 - \delta_{\tau, \tau'}$ which implies $\hat h(\omega) = t \delta_{\omega, 0} - 1$.
Numerical evidence supports the claim that in general $\hat h(\omega \neq 0) / \hat h(\omega = 0) \stackrel{t \to \infty}{\to} 0$. Under this assumption, one recovers the emergent \textit{Potts model} of SFF~\cite{supplementary} and the universal result from \cite{cdc2,chan2021manybody},
\be
\label{eq:kappafgin}
\kappafgin(x)  
= 
\lim_{\substack{L,t \to \infty\\ x =L/\Lsb(t) }}
   \kfgin
   -  t
   =
  e^{x} -x  - 1 
  \; ,
\ee
with $\Lsb(t) = [f(t) \hat h(0)]^{-1}$. Hence, we have for {case b}
\be \label{eq:floq_conj}
  \kappaginueB(x) 
  =\kappa_{\mathrm{F}-\mathrm{MBQC}} (x)  
\;.
\ee

\paragraph {Translation invariant Floquet case.}-- 
For TI Floquet systems {(case d)}, we model the transfer matrix with \eqref{eq:gin_corr}, except  that  TI is imposed, i.e. $\VG(t,r) = \VG(t,r')$ for all $r, r'$. In practice, Eq.~\eqref{eq:gin_corr} implies that different frequency sectors are statistically decoupled. We can thus evaluate $\ktifgin$ for this model, using Eq.~\eqref{eq:kginue} within each sector and replacing the variance $v/N \to \hat J(\omega)/\Nb$. Using the results in Eqs.~(\ref{eq:kc_scaling_dgt1},\ref{eq:kappafgin}), one obtains for $L\neq 0$,
\begin{equation}\label{eq:tif_k}
\begin{split}
\ktifgin (t,L) 
= 
\KGinUE (\Nb, L) \kfgin(t,L)  
\\
\sim L \, \kappaginueC \left(\frac{L}{\Lsc}\right) \left[\kappaginueB \left(\frac{L}{\Lsb}\right) + t \right] \;,
\end{split}
\end{equation}
and sees that the emerging scaling form depends on 
the ratio between the relevant length scales, namely $\Lsb$ and $\Lsc$.
%
For instance, if $\Lsc \ll \Lsb$ at large $t$, 
the appropriate scaling limit has $x= L/\Lsc$ fixed, giving the scaling form
\be\label{eq:caseD_collapse1}
\kappa^{(\mathrm{TI})}_{\mathrm{TIF}-\mathrm{MBQC}} (x) :=
 \lim_{\substack{L,t \to \infty\\ x =L/\Lsc}} 
 \frac{\ktifgin 
  }{
  tL
  }
  =   \kappa_{\mathrm{TI}-\mathrm{Gin}}(x) 
\ee
On the contrary, if $\Lsb \ll \Lsc$ at large $t$, the appropriate scaling limit has $x=L/\Lsb$ fixed leading to
\be\label{eq:caseD_collapse2}
\kappa^{(\mathrm{F})}_{\mathrm{TIF}-\mathrm{MBQC}} (x)  := 
 \lim_{\substack{L,t \to \infty\\ x =L/\Lsb}}  \frac{
 K_{\mathrm{TIF}-\mathrm{MBQC}}
 }{
 L
 }-t
  =   \kappaginueB(x)
\ee
To test this, in  Fig.~\ref{fig:three_panels}e, we simulate the TI Floquet brick wall model as a representative example, and show that an excellent fit can be obtained using Eq.~\eqref{eq:tif_k}, with $\Lsb$ and $\Lsc$ as fitting parameters in Fig.~\ref{fig:three_panels}f.
While we cannot determine the large-$t$ behaviour of $\Lsc$ $\Lsb$ from the finite size data, we can extrapolate that $\Lsc \ll \Lsb$ for this model, and obtain a consistent scaling collapse of \eqref{eq:caseD_collapse1} in \cite{supplementary}.
%
%
%
%
%
%
%


%
%
%
\textit{Discussion.}  The emergence of universal Ginibre behaviour complements the known emergence of Gaussian unitary ensemble in such systems, and opens up a new avenue to characterize quantum chaos. 
We emphasize that the emergence of GinUE is a \textit{many-body} quantum phenomenon: 
Firstly, the construction of spacetime duality requires spatial structure. 
Secondly, the crossover between linear ramp to exponential behaviours around $\Lth$ (or $\tth$) and the scaling collapse in the scaling limit is a manifestation of \textit{many-body} quantum effect --- the (connected) SFF of Gaussian and Circular ensembles have no exponential regime at all. 

\paragraph{Acknowledgements.}
We thank Vir Bulchandani, Soonwon Choi, Giorgio Cipolloni, Ceren Da\v{g}, Michael Gullans, Jonah Kudler-Flam, Igor Klebanov, Daniel Mark, Simeon Mistakidis, Adam Nahum, Vladimir Narovlansky, Hossein Sadeghpour, Grace Sommers and Tianci Zhou for feedback and discussions.
AC and ADL warmly thank John Chalker for his guidance in related projects.  DAH is supported in part by NSF QLCI grant OMA-2120757.
AC is supported by fellowships from the Croucher foundation and the PCTS at Princeton University. The numerics were performed using Princeton Research Computing resources at Princeton University. ADL acknowledges support by the ANR JCJC grant ANR-21-CE47-0003 (TamEnt).
%






\bibliography{biblio.bib}


\onecolumngrid
\newpage 

\appendix
\setcounter{equation}{0}
\setcounter{figure}{0}
\renewcommand{\thetable}{S\arabic{table}}
\renewcommand{\theequation}{S\thesection.\arabic{equation}}
\renewcommand{\thefigure}{S\arabic{figure}}
\setcounter{secnumdepth}{2}

\begin{center}
{\Large Supplementary Material \\ 
\vspace{0.2cm}
\titleinfo
\\
\vspace{0.33cm}
{\large
Saumya Shivam, Andrea De Luca, David A. Huse, and Amos Chan
}
}
\end{center}

In this supplementary material we provide additional details about:
\begin{enumerate}[label=\Alph*)]

    \item[A. ] Models of MBQC
    \begin{itemize}
        \item[1. ] Two-site random unitary gate
        \item[2. ] Random phase model (RPM)
        \item[3. ] Brick-wall model (BWM)
        \item[4. ] Kicked Ising model (KIM)
    \end{itemize}
    \item[B. ] Ginibre models of dual transfer matrices (DTM) of MBQC systems
    \item[C. ] SFF regime diagrams for MBQC systems with space-time translational invariance
    \item[D. ] Spectral properties of dual spectra 
        \begin{itemize}
        \item[1. ] Distribution of dual eigenvalues and unfolding
        \item[2. ] Spectra of single realization
        \item[3. ] Dual spectra of non-interacting quantum circuits
    \end{itemize}  
    \item[E. ] Spectral correlation of dual spectra
            \begin{itemize}
        \item[1. ]  Nearest neighbour spacing distribution
        \item[2. ] Complex spacing ratio
        \item[3. ] Dissipative spectral form factor (DSFF)
    \end{itemize}    

    \item[F. ] SFF for GinUE (TI GinUE model)
        \begin{itemize}
        \item[1. ]  Exact evaluation of SFF for GinUE
        \item[2. ] Diagrammatical approach
    \end{itemize}    
    
    \item[G. ] SFF for Floquet GinUE model
    \begin{itemize}
        \item[1. ] Long-range $J(\tau)$
        \item[2. ] Long-range $J(\tau)$ example: Emergent Potts model
        \item[3. ] Short-range $J(\tau)$
        \item[4. ] Short-range $J(\tau)$ example with nearest-neighbour coupling
    \end{itemize}
    \item[H. ] SFF for TI Floquet GinUE model
    \item[I. ] SFF: Numerical results
\end{enumerate}


\section{Models of MBQC}
In this appendix, we explicitly write down the models studied in the main text, namely, the random phase model (RPM), the brick wall model (BWM), and the kicked ising model (KIM). 
Additionally, we consider an ensemble of two-site random unitary gates, which are the building blocks of random quantum circuits, and are the simplest objects that admit the space-time duality. 
The time evolution of a translational invariant circuit with  $q$ degrees of freedom per site can be defined by the operator $\mathcal{W}(t,L)$, which is constructed from layers of the circuit $W(t,L)$, such that the time evolution operator can be written as 
\begin{align}
    \mathcal{W}(t,L)=\prod_{t'=1}^{t'=t} W(t',L) \;. 
\end{align}
 For the  case when the circuit is also periodic in time (Floquet or TIF models), we have $W(t',L) = W(L)$ for all values of $t'$, i.e. the time evolution operator can be written as
\begin{align}
    \mathcal{W}(t,L)=W(L)^t \;. 
\end{align}
The duality transformation described in Fig.~\ref{Fig:sff_reg} results in a dual evolution operator $\mathcal{V}(t,L)$, which for both TI and TIF models are composed of DTM $V(t,r) = V(t)$ for all values of $r$ (Fig.~\ref{Fig:sff_reg} red) according to
\begin{align}
    \mathcal{V}(t,L)=V(t)^L \;.
\end{align}
Note that this implies that $t,L$ denote the number of time $W$ and $V$ are multiplied, respectively, in order to obtain $\mathcal{W}$ and $\mathcal{V}$. The number of sites for either operator is related to $t,L$ but depends on the model details so we provide the explicit relation for each model in the following sections. With this general notation in mind, we now describe $\mathcal{V}(t,L)$ for a two-site unitary gate and three circuit models known to show chaotic behavior, namely the RPM, the BWM, and the KIM. 
%

\subsection{Two-site random unitary gate}\label{app:mod_2site}

\begin{figure}[H]
\centering
\includegraphics[width=0.25\textwidth]{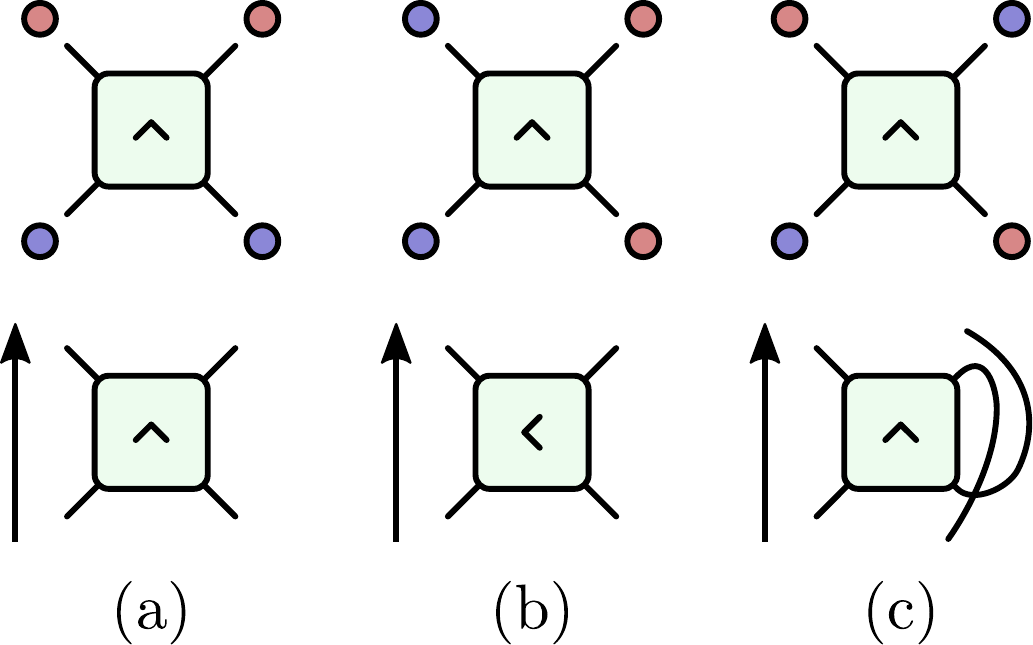}
\caption{
Top: Three ways of partitioning the four legs of a two-site gate into a pair of incoming (blue) and a pair of outgoing (red) tensor legs:
(a) a two-site gate acting in the time direction; 
(b) a two-site gate acting in the space direction;  
(c) a two-site gate with the bottom left and top right legs as incoming legs and the rest of the legs as outgoing legs. 
Bottom: An alternative representation of these two-site gates, with incoming legs in the bottom and outgoing legs at the top. 
}
\label{fig:twogates}
\end{figure}

A single two-site Haar-random unitary gate $u_{ab}^{cd}$ (Fig.~\ref{Fig:sff_reg} green), having dimensions $q^2 \times q^2$, serves as a simple model to study the correlations of the dual spectrum. 
In our notation, the two lower indices and upper two indices of $u_{ab}^{cd}$ label the columns and rows of the matrix $u$ respectively.
The corresponding dual gate $v$ is defined as $v_{ca}^{db}=u_{ab}^{cd}$, where the permutation of indices correspond to space-time rotation of the two-gate. 
%

In fact, given any two-site unitary gate (Fig.~\ref{fig:twogates}) which is a tensor with four legs, there are three ways to pair-wise partition the incoming and outgoing legs, namely (a)  two-site gate acting in the time direction (Fig.~\ref{fig:twogates}a); (b) two-site gate acting in the space direction (Fig.~\ref{fig:twogates}b); (c)  two-site gate acting in the time direction after a partial transpose (Fig.~\ref{fig:twogates}c). 
If the two-site unitary gate is drawn from the CUE (i.e. it is Haar-random), the ensemble (c) is equivalent to (b) due to the property of CUE being invariant under a unitary transformation. More explicitly, (c) can be obtained as the  space-time dual of (a) after applying an additional swap gate, but since the swap gate can in fact be absorbed into the CUE, (c) is in fact equivalent to (b) for the CUE.
In short, this paper studies the spectral correlation of (b) and (c) for a two-site Haar-random unitary gate, and show that they fall under the universality class of GinUE.

\subsection{Random phase model (RPM)}

The random phase model was first introduced as an analytically-tractable minimal model of many body quantum chaos without any symmetries ~\cite{chan2019}, with the Floquet operator $W(L)$ consists of layers of $W(L)=W_2(L) W_1(L)$, 
where the layer $W_1$ consists of on-site Haar random unitaries $u$ of dimension $q \times q$, and the layer $W_2$ consists of two-site diagonal phase gates 
with the phases chosen from a normal distribution with zero mean and variance $\epsilon$. 
Specifically, taking $\bv = (b_1, \ldots, b_t)$ with $b_\mu= 1,\ldots,q$ to define the computational basis, we write
 \begin{subequations}
\label{eq:VVsdef_bwm}
\begin{align}
&W_{1}(t)  = \prod_{r=1 }^L  u(t,r)  \;, \\
&[W_{2}]_{\bv,\bv'}(t)  =  \exp \left(  \imath  \sum_{r=1}^L 
\varphi_{b_{r}, b_{r+1}  }(t,r)
\right) \delta_{\bv,\bv'} 
\;.
\end{align}
\end{subequations}
such that $\mathcal{W}(t,L)$ is a $q^L \times q^L$ operator, and the number of sites is $L$. The TI version of the circuit, where the $r$-dependence above is removed, was recently also shown to display chaotic behavior~\cite{chan2021manybody}. 
The layers of $W(t,L)$ can be random (TI-RPM) or repeating (TIF-RPM).
Applying the duality transformation from Figure~\ref{Fig:sff_reg} on each gate leads to the dual of the single-site gate becoming a diagonal two-site gate and vice-versa for the original two-site diagonal gate. 
The DTM $V(t)$ for both TI-RPM and TIF-RPM, can be constructed in the following way using the two layers $V_1(t),V_2(t)$ such that $V(t)=V_2(t) V_1(t)$, where
\begin{subequations}
\label{eq:VVsdef}
\begin{align}
&[\VV_{1}]_{\bv,\bv'} (t) = 
\exp \left(  \imath  \sum_{t'=1}^t 
\varphi_{b_{t'}, b'_{t'}  }(t')
\right) 
\label{eq:VV1}  \;, \\
&[\VV_{2}]_{\bv,\bv'} (t) =  \prod_{t'=1}^t u^{b_{t'+1}}_{b_{t'}}(t') \delta_{\bv,\bv'} 
\;,
\end{align}
\end{subequations}
and where periodic boundary condition has been assumed. 
Note that the DTM $V(t)$ inherits the same geometry as the original circuit. This also implies that $\mathcal{V}(t,L)$ is a $q^t \times q^t$ operator, and the number of corresponding sites is $t$.
For the case of TIF-RPM, there are $t$ momentum sectors with each block having the approximate size $q^t/t$. 
We will restrict ourselves to the zero momentum sector, which turns out to have the largest size, for the remainder of the discussion. Moreover, we only consider $q=3$, as the circuits are not fully chaotic behavior for $q=2$~\cite{cdc2}.

\subsection{Brick-wall model (BWM)}
 The brick-wall model is constructed with two-site $q^2 \times q^2$ Haar-random unitaries $u_{ij}^{kl}(t',r,i)$ in the brick work geometry (Fig.~\ref{Fig:sff_reg}), 
 Specifically, we have bi-layer $W(t',L) = W_2(t',L) W_1(t',L)$ consisting of two layers $W_1$ and $W_2$ defined by
 \begin{equation}
\label{eq:VVsdef_bwm}
W_i(t',L) = T_{2L}^{\delta_{i,2}}
\left[ \bigotimes_{r \in L} u(t',r,i) \right] T_{2L}^{-\delta_{i,2}} \;,
\end{equation}
where, again, $T_{n}$ is the translational operator in the Hilbert space, $\bigotimes_{1}^n \mathbb{C}^q$, defined by its action on the computational basis, $T_{n} \ket{\mathbf{s} \equiv s_1, s_2, \dots, s_{n}} = \ket{T(\mathbf{s}) \equiv s_{n}, s_1, \dots, s_{n-1}}$.
This means that the circuit has $2L$ sites and therefore $\mathcal{W}$ acts on a $q^{2L}$ dimensional Hilbert space. 
The dual Floquet operator $V(t)=V_2(t) \, V_1(t)$, for TI-BWM, is then described by the two layers $\VV_1(t)$ and $\VV_2(t)$, such that in the computational basis  
\begin{equation}
\label{eq:VVsdef_bwm}
V_i(t,r) = T^{\delta_{2,i}}_{2t} \left[\bigotimes_{t'=1}^t v(t' ,r, i) \right] T^{-\delta_{2,i}}_{2t}
\;,
\end{equation}
where $v(t',r,i)$ is the dual of the unitary gate $u(t',r,i)$ (with TI assumed without the $r$-dependence), defined by $v_{ca}^{db}=u_{ab}^{cd}$ (see Fig.~\ref{Fig:sff_reg}), such that $\mathcal{V}$ acts on a $q^{2t}$ dimensional Hilbert space.
%
%
%

%
Periodic boundary conditions have again been assumed. 
Note that for TIF-BWM, the unit cell in the time direction contains two sites, leading to $t$ momentum sectors, and we will restrict ourselves to the zero momentum sector.

\subsection{Kicked Ising model (KIM)}
One of the ways in which duality of quantum circuits has been utilised is by studying a self-dual model, which arises at special points in the parameter space of the kicked Ising model, and self duality imparts analytical tractability to proving emergence of random matrix theory through exact derivation of quantities like the spectral form factor\cite{bertini_exact_2018,bertini_exact_2019,bertini_op_2020}. The model has the same geometry as the random phase model, and the time evolution operator (acting on $q^L$ dimensional Hilbert space) 
are defined using the two layers 
\begin{align}
\begin{split}
& W_{1} (t',L) =\prod_{r=1}^L  e^{\imath h(r,t')\,  \sigma_{r}^z}e^{ \imath b \, \sigma^x_{r}} \\
& W_{2}  (t',L) = \prod_{r=1}^L e^{\imath J \, \sigma^z_{r}{\sigma^z_{r+1}} }
\end{split}
\end{align}
where $J$ and $b$ are constants, while $h(r,t)$ is drawn from a normal distribution with variance $\sigma^2$. 
Note that the second layer $W_2(t',L)$ is diagonal. 
%
The DTM (acting on $q^t$ dimensional Hilbert space)  is then given by the layers 
\begin{align}\label{eq:kim_DTM}
\begin{split}
    &[\VV_{1}]_{\bv,\bv'} (t,r) = \prod_{\mu=1}^t e^{\imath J b_{\mu}b'_{\mu} }\\
    &[\VV_{2}]_{\bv,\bv'} (t,r) =\prod_{\mu=1}^t  e^{\imath h(r,\mu) \,  b_{\mu}}R_{b_{\mu},b_{\mu+1}} ~ \delta_{\bv,\bv'} 
\end{split}
\end{align}
so that the DTM is given by $V(t,r)=\VV_{2}(t,r) \VV_{1}(t,r) $, and $R_{1,1}=R_{-1,-1}=\cos(b)$ and $R_{1,-1}=R_{1,-1}=-\imath \sin(b)$. 
At $|J|=|b|=\pi/4$, the dual circuit $V$ becomes unitary 
which is the origin of the term ``dual-unitary'' for such circuits. The TI version of the model, TI-KIM, is obtained by removing the $r$-dependence in $h(r,t)$.

We study 
the dual spectra as we moves away from the self dual point, by decreasing $J$ from $\pi/4$ to $0$, while keeping $b=\pi/4$ fixed. 
When translational symmetry is not present, the dynamics undergo a transition when the interaction strength is reduced away from the self dual point~\cite{braun_transition_2020}. However, for TI circuits ($h_i$ same for each site), we don't expect the onset of many body localization, rather a transition to the non interacting integrable point at $J=0$.

Owing to the time reversal symmetry of the original circuit, the dual circuit also has a generalized time reversal symmetry, such that there exists a unitary operator $C_{+}$ such that $C_{+}V^TC_{+}^{-1}=V$ and $C_{+}C^*_{+}=\mathbb{1}$. Here $C_{+}$ can be chosen to be $2^{t/2} V_2^*$ for $b=\pi/4$. Hence, instead of GinUE, the corresponding class $\mathrm{AI}^{\dag}$ \cite{hamazaki_universality_2020} (generated by choosing the complex normal variables for each matrix element such that the matrix is symmetric) is more apt for comparison. Given the lack of detailed analytic understanding of this class of non-Hermitian random matrices, we will only evaluate and compare the level spacing distribution, and look at how the dual spectrum changes when moving away from the self dual point.

\section{Ginibre models of dual transfer matrices (DTM) of MBQC systems}
\begin{figure}[H]
\centering
\includegraphics[width=0.95\textwidth]{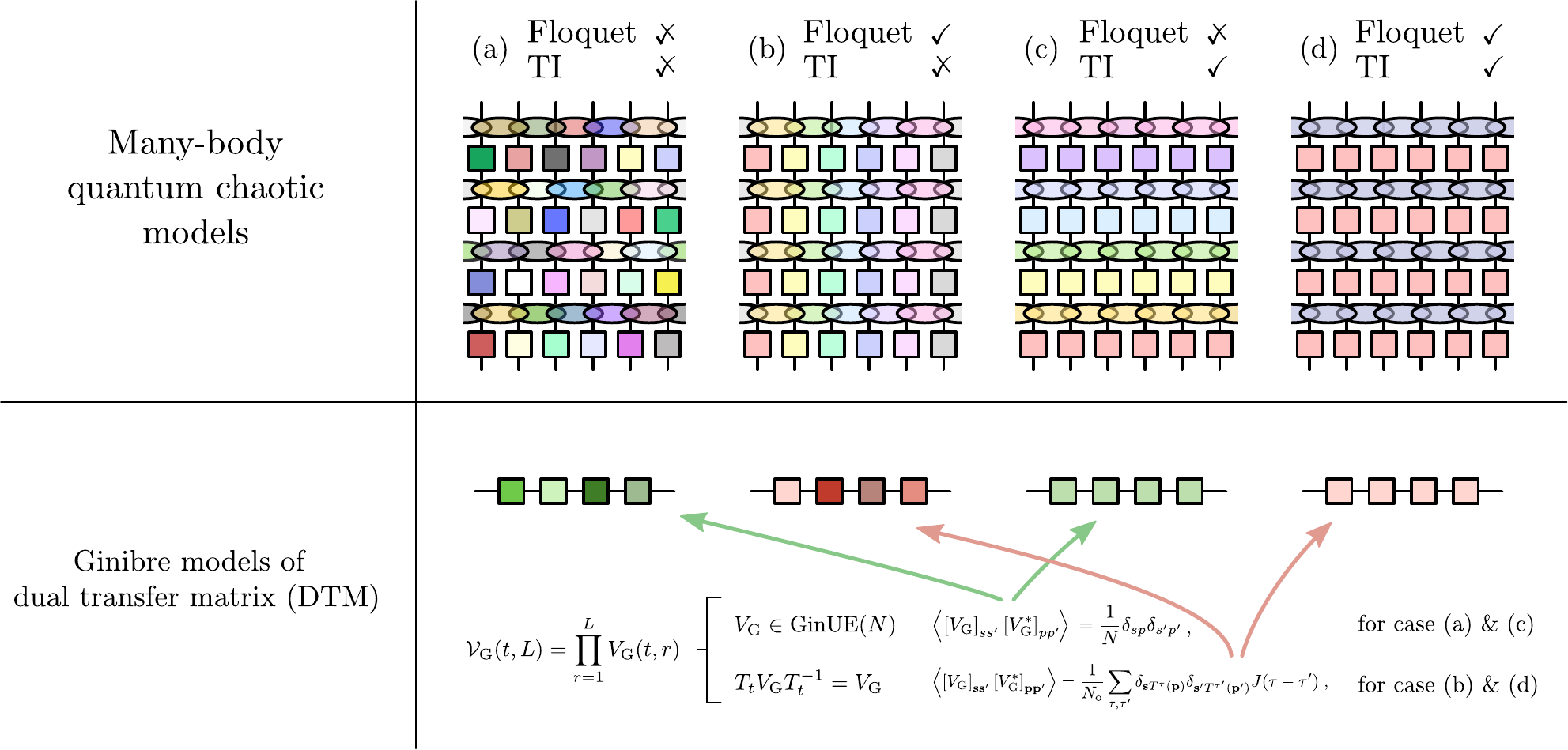}
\caption{
Summary of variations of Ginibre ensembles as model of DTM of MBQC systems.
}
\label{fig:gin_model}
\end{figure}

%
In this appendix, we provide a summary of the variations of Ginibre ensembles as models of dual transfer matrix (DTM) of  MBQC systems with translational symmetries in space and in time in Fig.~\ref{fig:gin_model}.
For systems without any symmetries, denoted as case (a) in Fig.~\ref{fig:gin_model}, 
%
the DTM of the MBQC systems can be modelled as a sequence of  matrices independently-drawn from the Ginibre ensemble, and its SFF can be computed trivially as 1 for all $L$, as consistent with the calculation of the MBQC circuit \cite{chan2021manybody}.
For Floquet systems without TI (case (b) in Fig.~\ref{fig:gin_model}), 
the DTM of MBQC systems is modelled by a sequence of independently-drawn matrices satisfying Eq. \eqref{eq:gin_corr}.
For TI systems with temporal randomness (case (c) in Fig.~\ref{fig:gin_model}), 
the DTM of MBQC systems is modelled by the repeated action of a matrix drawn from the Ginibre ensemble.
Lastly, for TI Floquet systems (case (d) in Fig.~\ref{fig:gin_model}), 
the DTM of MBQC systems is modelled by the repeated action of a random matrix satisfying Eq. \eqref{eq:gin_corr} .

%
%

\section{SFF regime diagrams for MBQC systems with space-time translational invariance}
\begin{figure}[H]
\centering
\includegraphics[width=0.95\textwidth]{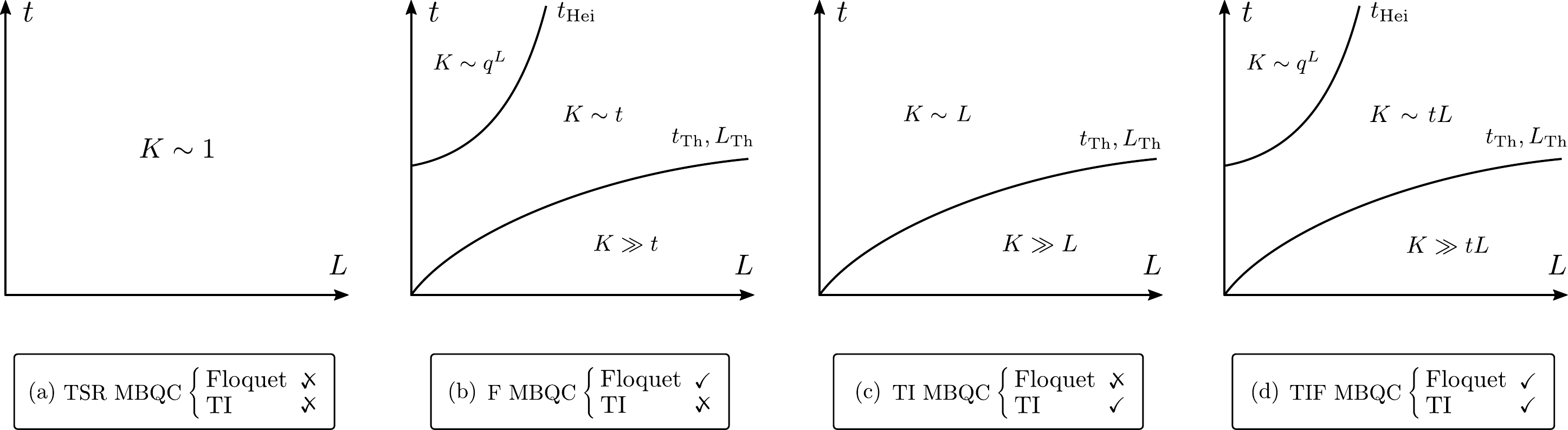}
\caption{
SFF regime diagrams for (a) temporal and spatial random (TSR) MBQC systems; (b) temporal periodicity, i.e. Floquet, and spatial random (Floquet) MBQC systems; (c) Floquet and translational invariant (TI) MBQC systems; and (d) Floquet and translational invariant (TIF) MBQC systems.
}
\label{fig:regime_diag_4cases}
\end{figure}

In this appendix, we present the regime diagrams for  MBQC systems with translational invariance in space and in time in Fig.~\ref{fig:regime_diag_4cases}.
Note that we focus on unitary models, such that the density of states is flat, i.e. without edges, and consequently there is not the so-called ``dip'' (see e.g. \cite{Gharibyan_2018}).
Note also that for temporal random MBQC systems, namely case a and c, the Heisenberg time $t_{\mathrm{Hei}}$ lines are absent.

\section{Spectral properties of dual spectra}
In this section, we show the distribution of eigenvalues and examples of what a typical dual spectrum looks like for the models defined in the previous section. Then, we discuss the unfolding procedure and the resulting eigenvalue distribution. 
%

\subsection{Distribution of dual eigenvalues and unfolding}\label{unfolding}
\begin{figure*}[t]
\begin{minipage}[t]{0.31\textwidth}
\includegraphics[width=1.1\linewidth,keepaspectratio=true]{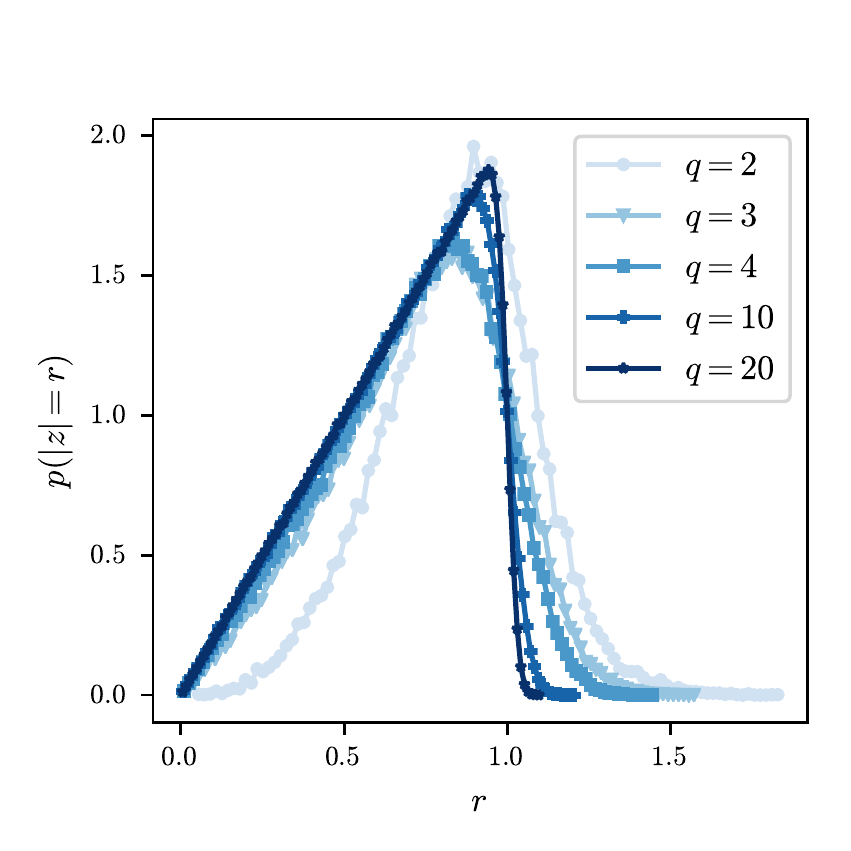}
\end{minipage}
\hspace*{\fill} 
\begin{minipage}[t]{0.31\textwidth}
\includegraphics[width=1.1\linewidth,keepaspectratio=true]{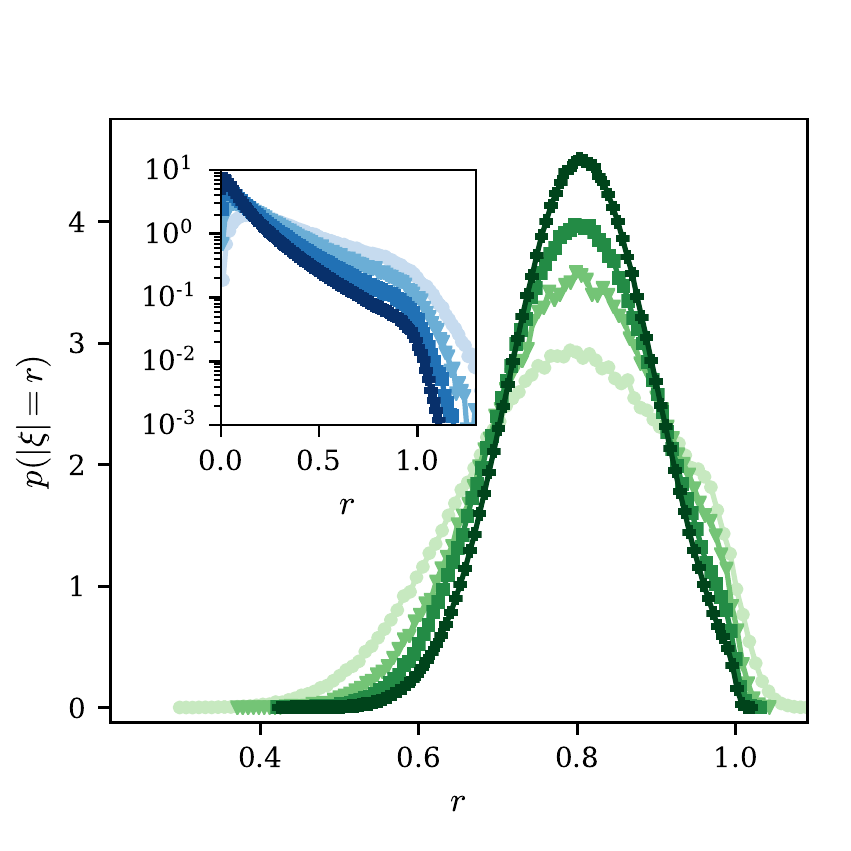}
\end{minipage}
\hspace*{\fill} 
\begin{minipage}[t]{0.31\textwidth}
\includegraphics[width=1.1\linewidth,keepaspectratio=true]{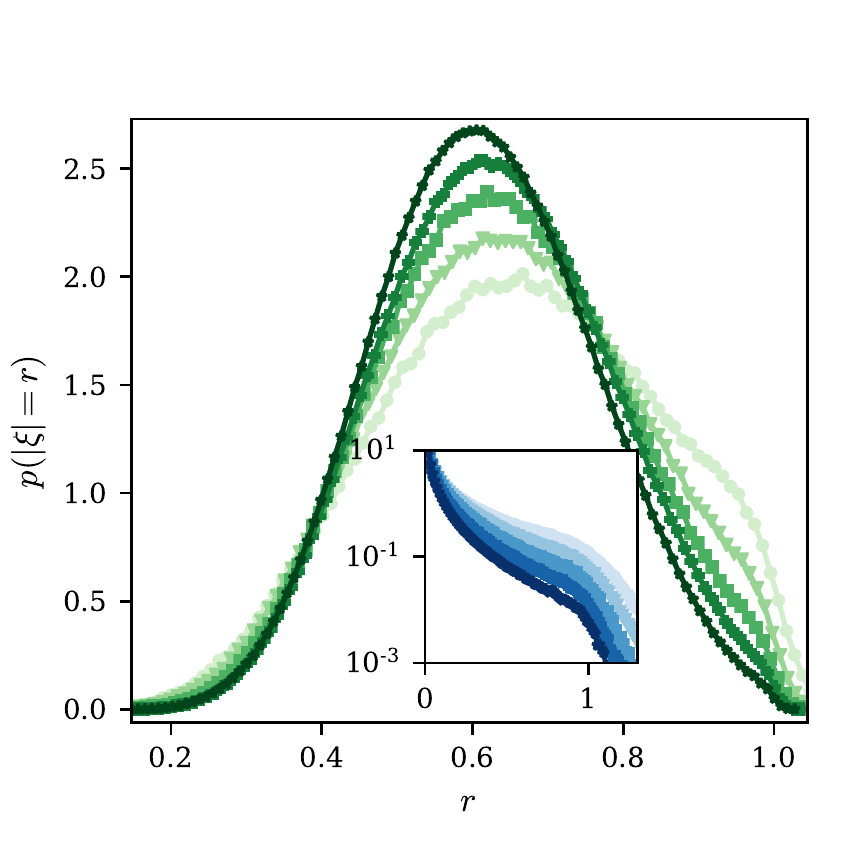}
\end{minipage}
\caption{Left : Radial distribution of dual eigenvalues of a two-site Haar-random unitary gate, with each site having degree of freedom $q$, for $q=2,3,4,10,20$ (increasing in darkness). 
Middle: Distribution of the magnitude of eigenvalues of TI-BWM with $q=2$ and $t=3,4,5,6$ with increasing darkness (inset), 
and the main panel shows the distribution of the magnitude of the unfolded spectrum $\xi=z^{\frac{1}{2t+1}}$.
Right: Distribution of the magnitude of eigenvalues of TI-RPM with $q=3$ and $t=4,5,6,7,8$ with increasing darkness (inset). The main panel shows the distribution of the magnitude of the unfolded spectrum $\xi=z^{\frac{1}{t+1}}$.
Using the slowly-varying regime of the unfolded spectrum, we can select of range $r_a < r <r_b$ so that for all system sizes, the distribution in that regime is approximately constant.
    } \label{fig:eigs_abs_dist}
\end{figure*}

In Figure \ref{fig:eigs_abs_dist}, the distribution of the absolute value of the eigenvalues is shown for the dual of a two-site unitary gate, TI-BWM and TI-RPM as representative examples, collected from around $1000-5000$ random realizations. 
For the two-site gate, for $q>2$, the distribution looks close to that of GinUE, which should be linear in the radial distance $r$ until $r\approx 1$. 
On the other hand, for the dual circuits, the distribution is heavily concentrated close to the origin. 
Note that it is sufficient to look at the radial distribution because the distribution of eigenvalues is rotational symmetric upon taking disorder average, which can also be seen in a scatter plot of the spectra of single realizations of such models in Figure \ref{fig:eigs_single}.

Since choosing a flat region from the original distribution for dual circuits can be difficult because of the power law/exponential distribution of the magnitude of eigenvalues, we take inspiration from the unfolding procedure performed in~\cite{akemann_universal_2012}. \cite{akemann_universal_2012} considers an ensemble of products of $n$ independent $N \times N$ matrices drawn from GinUE, and applies the conformal transformation $\xi = z^{1/n}$~\cite{akemann_universal_2012} to the spectrum, which reveals that the ensemble  eigenvalue distribution and correlations are GinUE-like. 
We apply the same transformation by associating $n$ with the number of (dual) sites or the number of gates in the dual circuit. 
Even though the circuit geometry is different from a simple product of matrices, the transformation leads to a more slowly varying distribution of eigenvalues (Figure \ref{fig:eigs_abs_dist}), allowing us to directly select a cutoff range of $r$ where using rectangular strips approximately leads to flat density. It remains to be seen if there exists a transformation which can make the distribution of eigenvalues completely flat.

\subsection{Spectra of single realizations}
\begin{figure*}[t]
\begin{minipage}[t]{0.31\textwidth}
\includegraphics[width=1.1\linewidth,keepaspectratio=true]{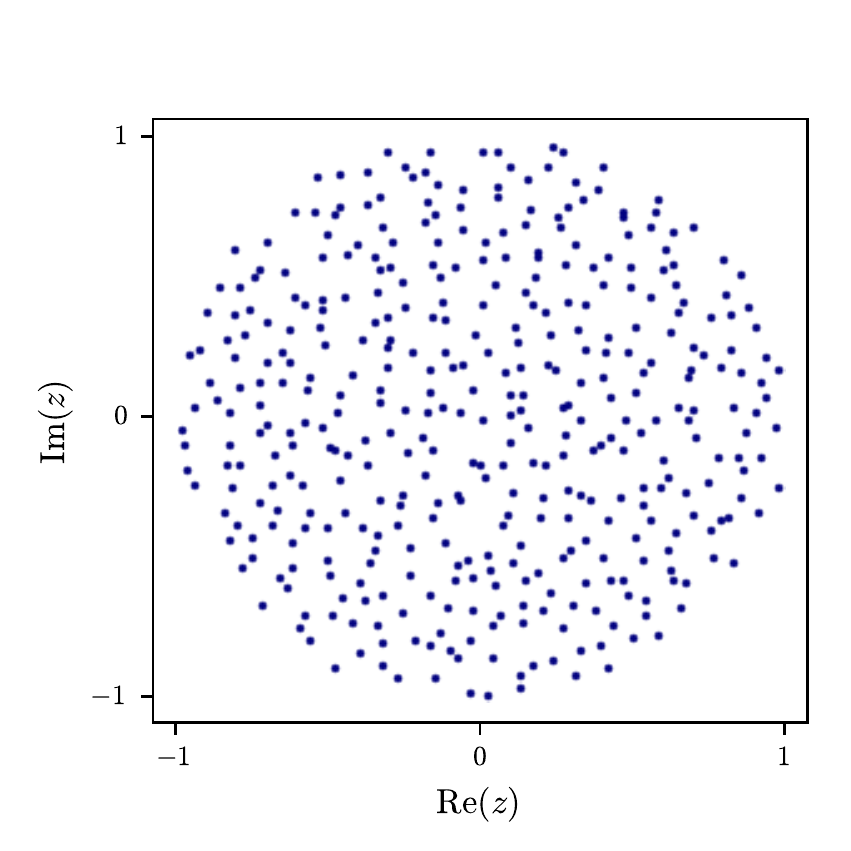}
\end{minipage}
\hspace*{\fill} 
\begin{minipage}[t]{0.31\textwidth}
\includegraphics[width=1.1\linewidth,keepaspectratio=true]{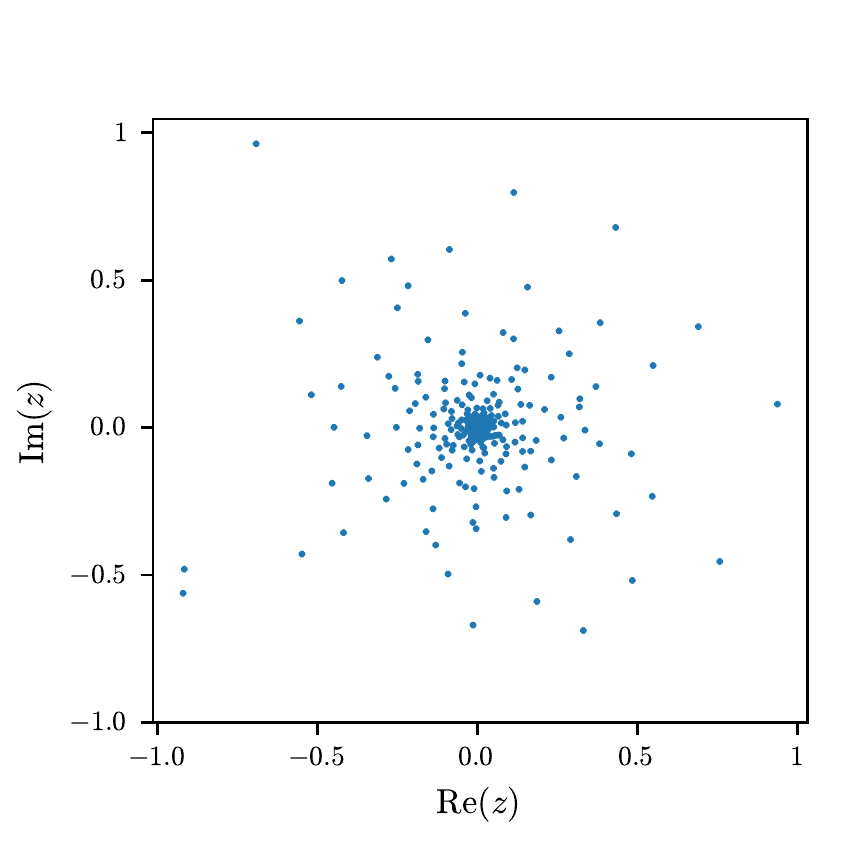}
\end{minipage}
\hspace*{\fill} 
\begin{minipage}[t]{0.31\textwidth}
\includegraphics[width=1.1\linewidth,keepaspectratio=true]{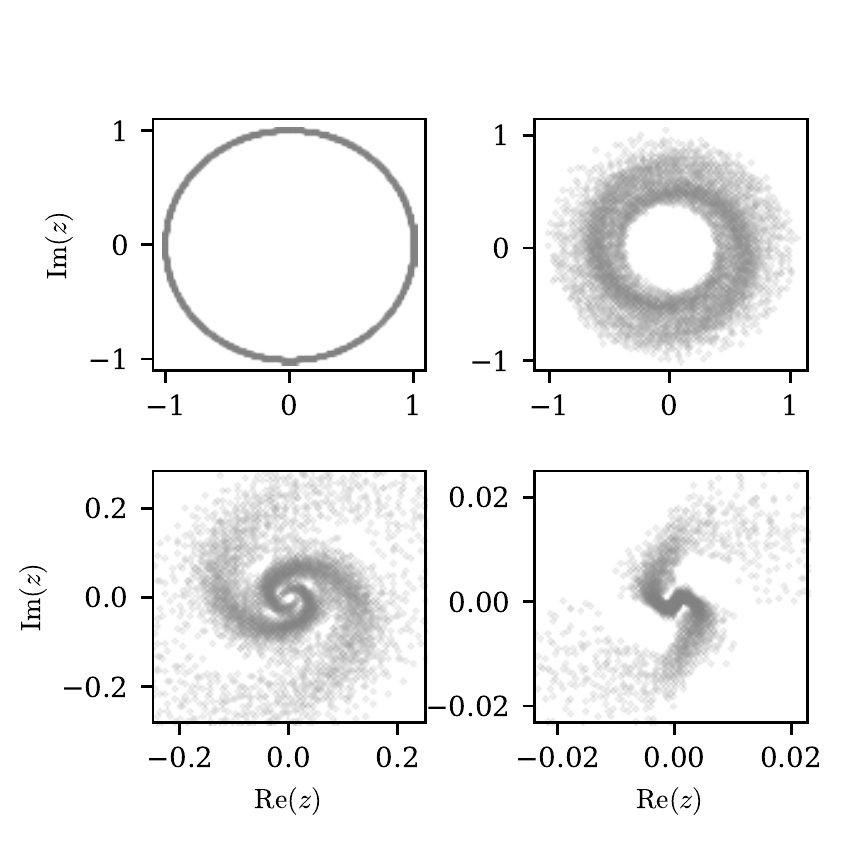}
\end{minipage}
\caption{Left: Spectrum of a single realization of the dual of a single two-site $q^2 \times q^2$ unitary with $q=20$. 
Middle: Scatter plot of the eigenvalues of single realizations of the DTM for the RPM, with $t=5$ and $q=3$. 
Right: Scatter plot of the eigenvalues of single realizations of the DTM for the KIM, with $t=12$ and $J/J_c=1,3/4,1/2,1/4$, clockwise from top-left, showing how starting from the self dual point, the eigenvalues collapse into a spiral, which has a fractal like structure that appears alongside level repulsion. 
    } \label{fig:eigs_single}
\end{figure*}

The spectrum of a single realizations of the dual of a two-site unitary gate, TI-RPM and a few parameters of KIM are shown in Figure \ref{fig:eigs_single}.
Note that for the dual two-site gate, the scatter plot looks uniformly spread out (with level repulsion) inside the unit disc. However for TI-BWM, they are more concentrated near the origin. Unlike both, however, TI-KIM displays a more complex structure of the spectrum. 
Starting with $J=\Jc$ (the self dual point), the spectrum is spread out on the unit circle, owing to the unitarity. But breaking the unitarity by decreasing $J/\Jc$ leads to the eigenvalues spreading inside the unit circle in a spiral structure, which is expected to continue exhibiting level repulsion, resulting in a fractal distribution.

\subsection{Dual spectra of non-interacting quantum circuits}

In this subsection, we comment on the dual spectra of non-interacting many-body quantum circuits.
The time direction quantum circuits are known to exhibit level statistics corresponding to the Poisson distribution when the quantum circuits are non-interacting or many-body localized~\cite{cdc1,ponte_floquet_mbl},
and we show here that such spectral properties are also reflected in dual spectra of the dual transfer matrix.
For concreteness, consider the TI-RPM (or TI-KIM) at the non-interacting point $\epsilon=0$ ($J=0$), where the contribution to the dual Floquet operator in Eq. \ref{eq:VVsdef} (Eq. \ref{eq:kim_DTM}) from the random phases becomes 
 \begin{equation}
 [\VV_{1}]_{\bv,\bv'} = 1
\end{equation}
i.e. $\VV_{1}$ becomes a rank-1 matrix with where all matrix elements are one, while $\VV_{2}$ remains a diagonal matrix. 
Consequently, the dual tensor $V = \VV_{2} \VV_{1}$ has the same elements for each column in a given row. 
Such a matrix has $N-1$ zero eigenvalues, and one non zero eigenvalue ($=\mathrm{Tr}[\VV_{2}]$).


\section{Spectral correlations of dual spectra}


Now we show the comparison of short and long range dual spectral statistics of various MBQC models and the GinUE, starting with nearest neighbour level spacing distribution, then the complex level spacing ratio, and finally the dissipative spectral form factor.


\subsection{Nearest neighbour spacing distribution}

\begin{figure*}[t]
\begin{minipage}[t]{0.31\textwidth}
\includegraphics[width=1.1\linewidth,keepaspectratio=true]{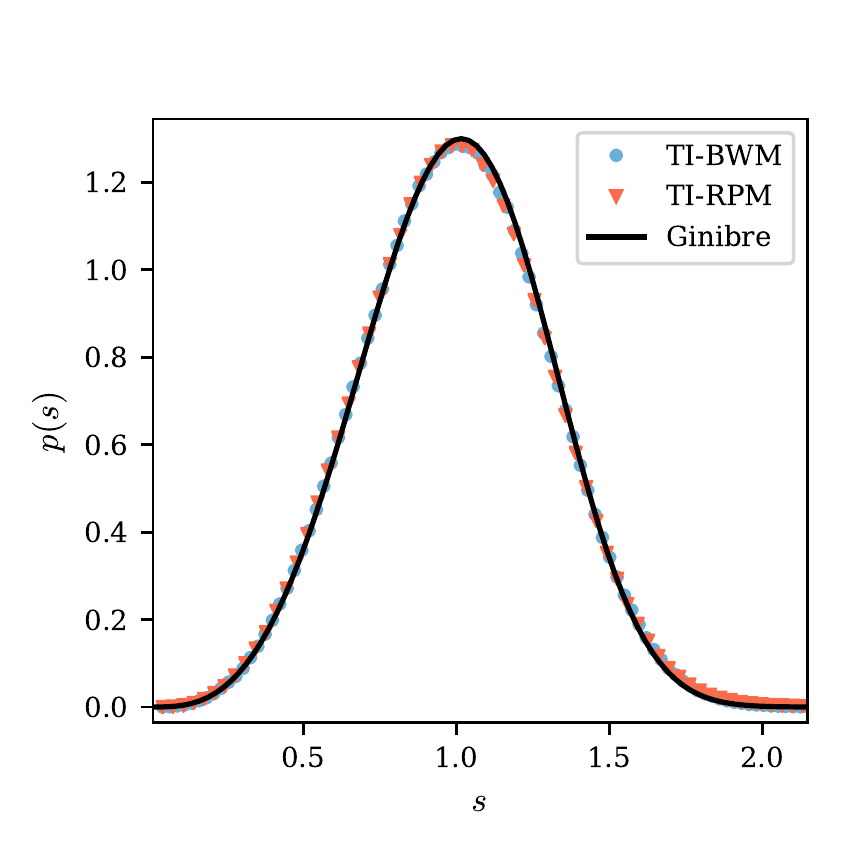}
\end{minipage}
\hspace*{\fill} 
\begin{minipage}[t]{0.31\textwidth}
\includegraphics[width=1.1\linewidth,keepaspectratio=true]{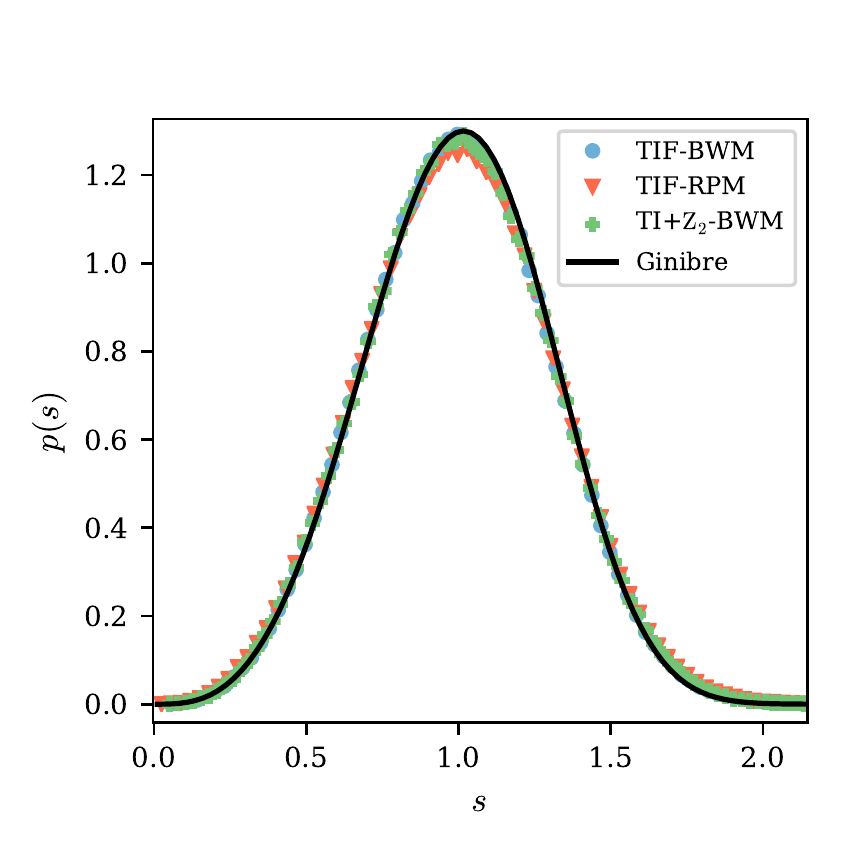}
\end{minipage}
\hspace*{\fill} 
\begin{minipage}[t]{0.31\textwidth}
\includegraphics[width=1.1\linewidth,keepaspectratio=true]{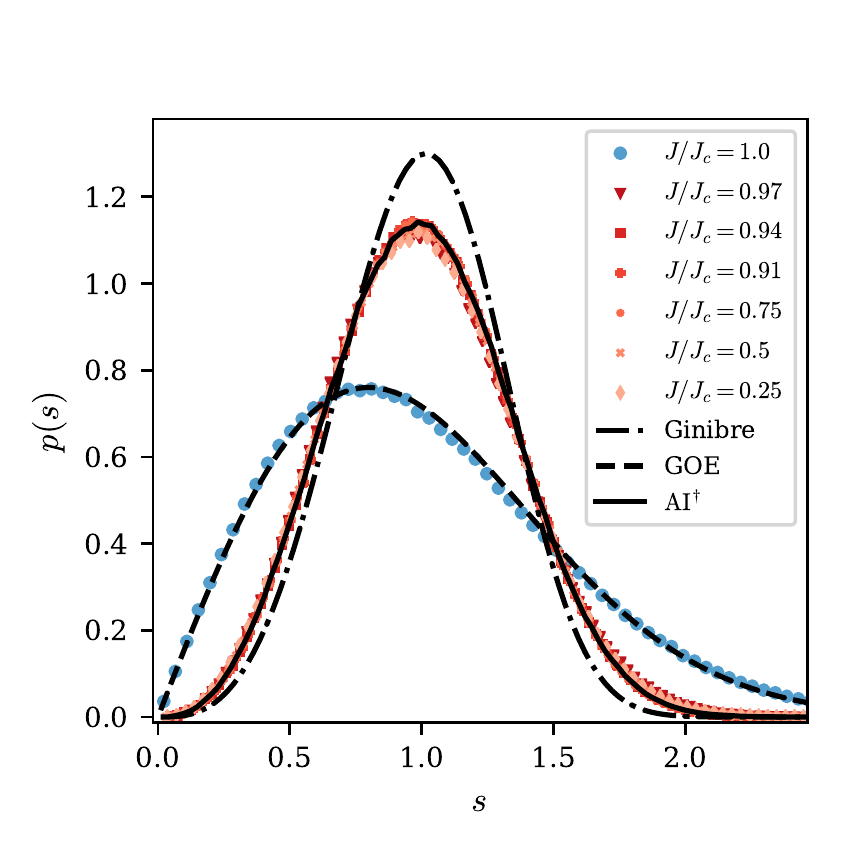}
\end{minipage}
\caption{Left: Unfolded and normalized level spacing distribution for dual of TI-RPM, with $q=3$, $\epsilon =2, t=8$ (red), and TI-BWM with $q=2,t=6$ (blue) compared with the exact distribution for the complex Ginibre ensemble (solid curve) showing good agreement. Middle: Unfolded and normalized level spacing distribution in a given symmetry sector for dual circuits with different types of symmetry, compared with the exact distribution for the complex Ginibre ensemble (solid curve) showing good agreement. 
Right: Unfolded and normalized level spacing distribution for dual of TI-KIM, with $b=\pi/4$, $\sigma =10\pi$ and $J$ varying from the value at self dual point $J=\pi/4$, whose dual spectrum falls under the universality class of GOE due to the time reversal symmetry. 
On the other hand as soon as we move away from the self dual point, the level spacing distribution resembles that of the time reversal symmetric Ginibre universality class $\mathrm{AI}^{\dag}$.
    } \label{fig:ls_ratio_dist}
\end{figure*}

The level spacing distribution of eigenvalues in the complex plane $p(s)$, can be computed by looking at the first and $n^{\mathrm{th}}$ nearest neighbour (in terms of the distance between the two points in the plane) for a given eigenvalue, denoted by $d_1$ and $d_n$. 
The unfolded level spacing is then given by $s=d_1\sqrt{n/\pi d_n^2}$~\cite{Haake}, and normalized such that the mean level spacing $\int s p(s)ds=1$. We choose $n$ in the range $[5,15]$, and plot the distribution of the entire spectrum over $1000$ to $5000$ realizations for different models in Figure \ref{fig:ls_ratio_dist}. Whenever an additional conserved quantity is incorporated, we restrict ourselves to a given subsector or block, for example for TIF circuits, we only look at the part of the spectrum belonging to the zero momentum sector or the sector with the largest dimensions.

For the KIM, as we move from $J=\pi/4$ to $J=0$ , the level spacing distribution, after performing the appropriate unfolding (with $n=15$ for $J<J_c$), changes
from that of the Gaussian Orthogonal Ensemble (GOE) ($J=J_c$) to $\mathrm{AI}^{\dag}$~\cite{hamazaki_universality_2020}, for which the analytical distribution is not known, so the distribution numerically generated from symmetrized GinUE matrices of size around $2000$ has been plotted for comparison. 
The distribution for $J=0$ is not plotted but can be shown to be the same as for TI-RPM with $\epsilon=0$, having $N-1$ eigenvalues zero, and a single non zero value. A summary of the 
universality classes for different models is given in Table \ref{tab:symm_class}.

\begin{table*}[h]
\centering
\begin{tabular}[t]{p{0.2\linewidth}>{\raggedright}p{0.2\linewidth}>{\raggedright\arraybackslash}p{0.2\linewidth}}
\toprule
    Space direction & Time direction  & Universality Class  \\
 \midrule
    Translation & None & GinUE  \\
    Translation & Translation  & GinUE  \\
    Translation + TRS & $\mathrm{TRS}^{\dag}$ &  $\mathrm{AI}^{\dag}$  \\
\bottomrule
\end{tabular}

    \caption{Different symmetric circuits considered here, with the first two columns noting the symmetry in the space and time direction, and the third column denoting the resulting symmetry in the dual spatial direction. The third column denotes the conjectured universality class of the corresponding circuit in a given projected sector of the conserved quantity corresponding to the symmetry of the dual circuit.}
    \label{tab:symm_class}
\end{table*}%

\subsection{Complex spacing ratio}
\begin{figure*}[t]
\begin{minipage}[t]{0.31\textwidth}
\includegraphics[width=1.1\linewidth,keepaspectratio=true]{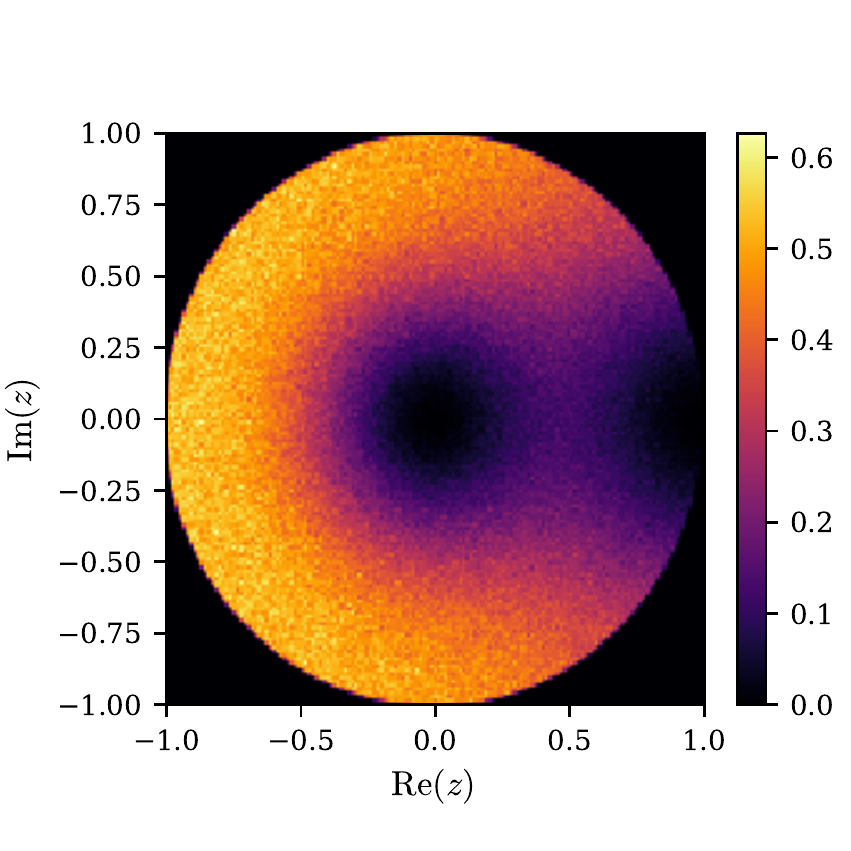}
\end{minipage}
\hspace*{\fill} 
\begin{minipage}[t]{0.31\textwidth}
\includegraphics[width=1.1\linewidth,keepaspectratio=true]{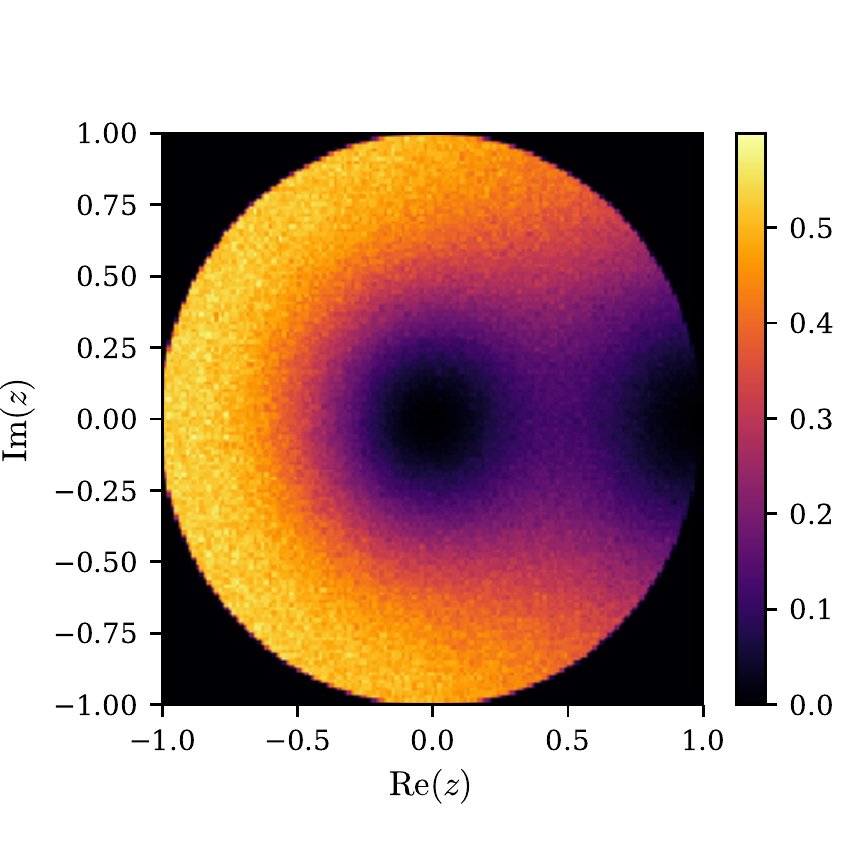}
\end{minipage}
\hspace*{\fill} 
\begin{minipage}[t]{0.31\textwidth}
\includegraphics[width=1.1\linewidth,keepaspectratio=true]{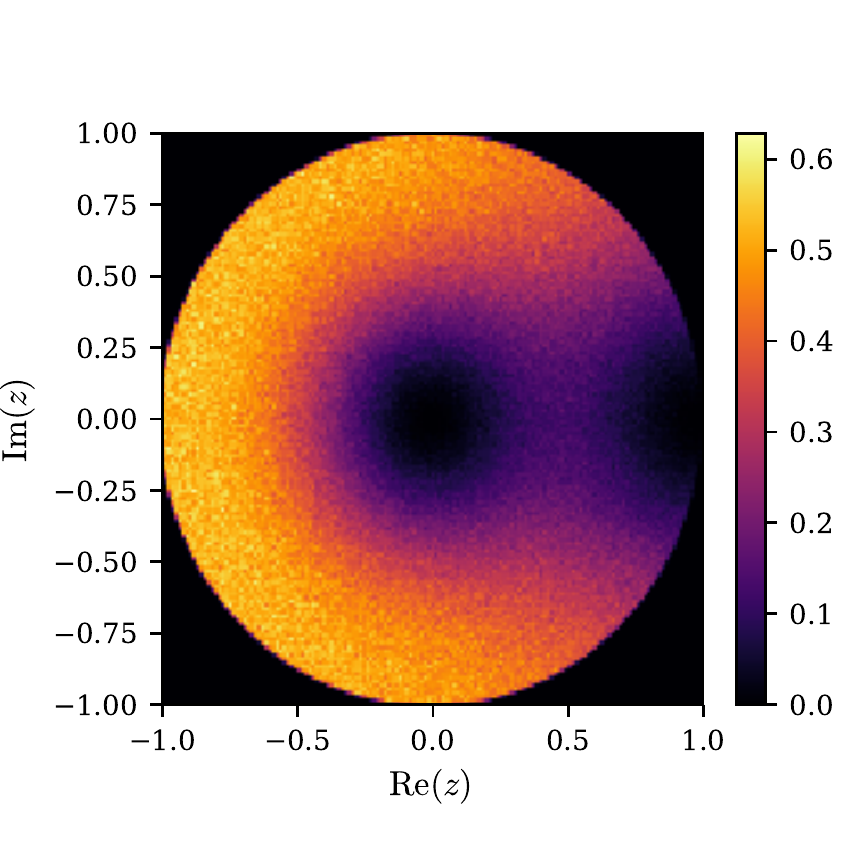}
\end{minipage}
\caption{Complex level spacing ratio distribution for GinUE (left), TI-RPM (middle) and TI-BWM (right), with the two dual spectra showing excellent qualitative agreement with the typical pac-man shape of level repulsion for the Ginibre ensemble. These results are consistent with the corresponding nearest neighbour level spacing distributions.
    } \label{fig:z_ratio_dist}
\end{figure*}

Another useful quantity which also probes the angular correlations at the shortest scale is the complex level spacing ratio $r$ \cite{sa_complex_2020}, given for the eigenvalue $z_i$ by
\begin{align}
    r_i = \frac{z_{i}^{\mathrm{nn}}-z_i}{z_{i}^{\mathrm{nnn}}-z_i} 
\end{align}
where $z_{i}^{\mathrm{nn}}$ and $z_{i}^{\mathrm{nnn}}$ are the nearest and next nearest distance neighbours of $z$ respectively. 
The distribution of $r$ in the complex plane has a ``pac-man'' shape for non-Hermitian matrices which show level repulsion \cite{sa_complex_2020,garcia-garcia_symmetry_2021}. 
An advantage of this quantity is that it doesn't require unfolding of the spectra, unlike the level spacing distribution. Figure \ref{fig:z_ratio_dist} shows qualitative resemblance of the distribution for TI-RPM and TI-BWM with that of the GinUE.

\subsection{Dissipative spectral form factor (DSFF)} 
\begin{figure*}[t]
\begin{minipage}[t]{0.31\textwidth}
\includegraphics[width=1.1\linewidth,keepaspectratio=true]{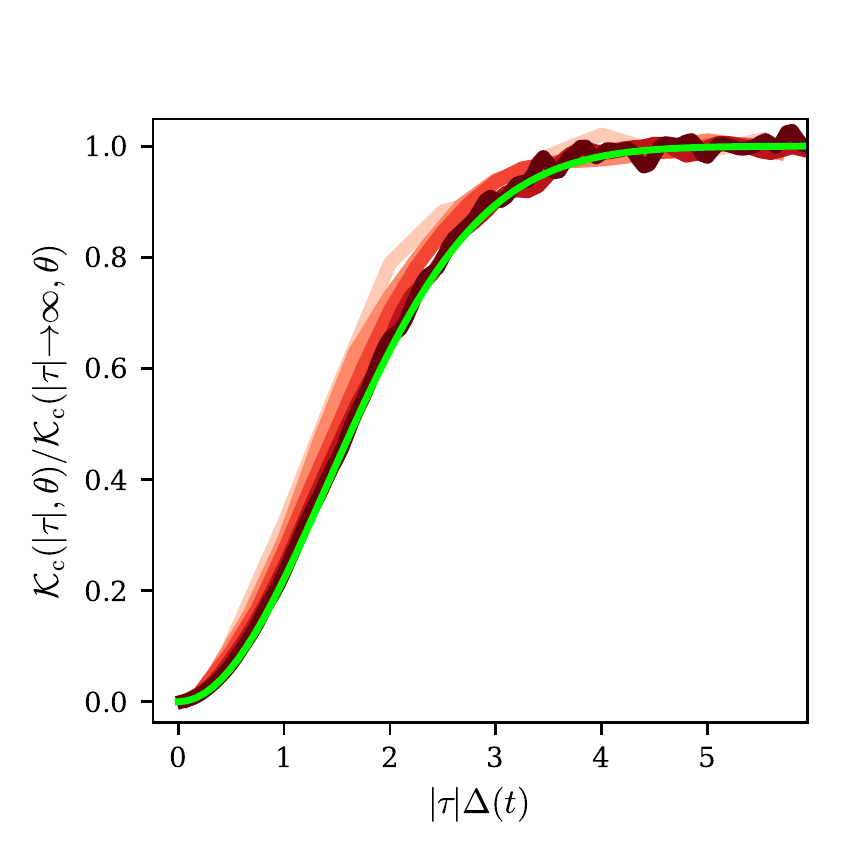}
\end{minipage}
\hspace*{\fill} 
\begin{minipage}[t]{0.31\textwidth}
\includegraphics[width=1.1\linewidth,keepaspectratio=true]{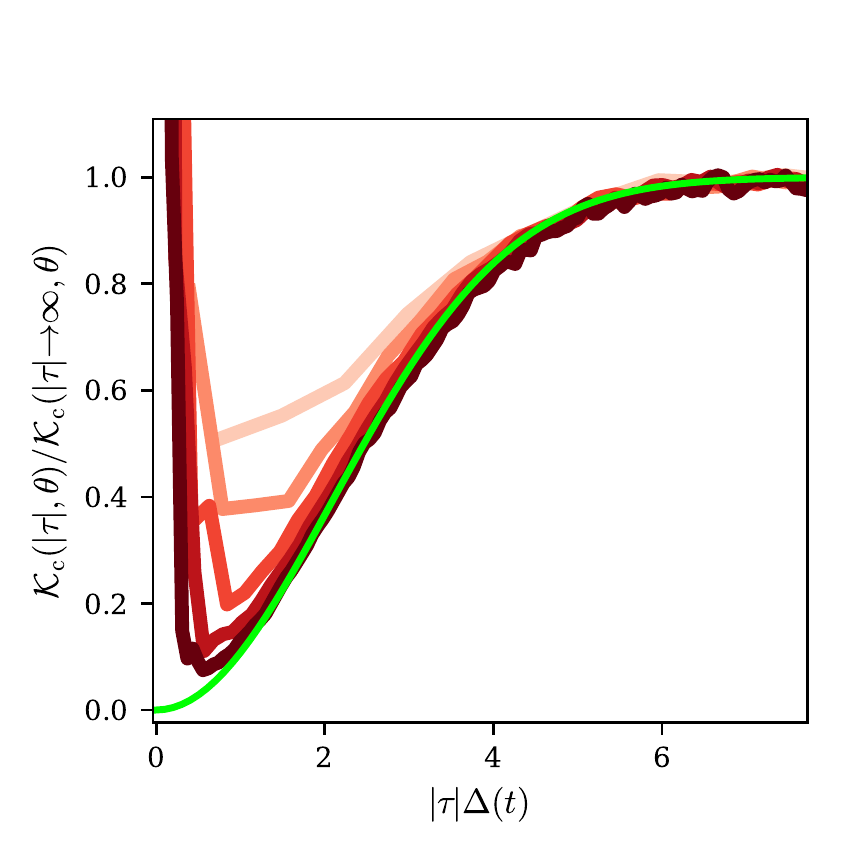}
\end{minipage}
\hspace*{\fill} 
\begin{minipage}[t]{0.31\textwidth}
\includegraphics[width=1.1\linewidth,keepaspectratio=true]{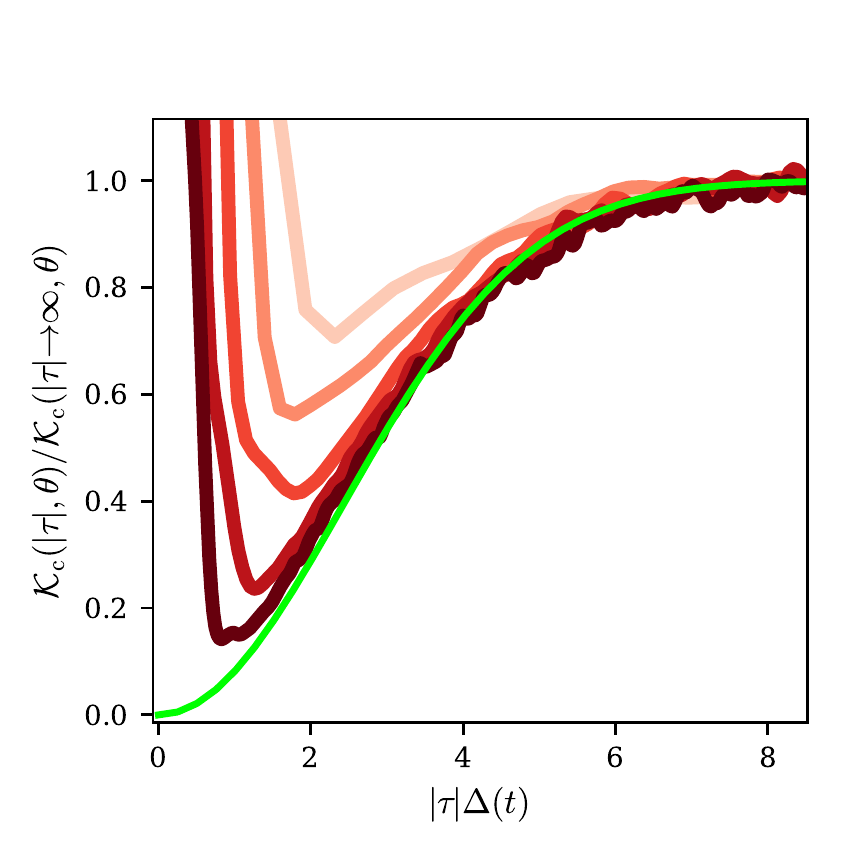}
\end{minipage}
\caption{Left: Connected part of the DSFF of dual of two-site $q^2 \times q^2$ Haar-random unitary gate with $\theta=\pi/4$ and $q=2,3,4,10,20$ and the darkness increasing for larger $q$.
  The  rescaled $\dsffcgin(|\tau|)$ is given as the green curve. 
  No unfolding is performed, resulting in finite size effects for small $q$, although for larger $q$ the spectrum is generally flat(Figure \ref{fig:eigs_abs_dist}). 
  %
%
Middle: Connected DSFF of TI-RPM with $\theta=\pi/8$, $q=3$, and $t=4,5,6,7,8$.
Right: Connected DSFF in the zero momentum sector of TIF-BWM with $q=2$, $\theta=\pi/16$, and  $t=6,8,10,12,14$. 
All three different models approach the GinUE behaviour as system sizes increase.
    } \label{fig:dsff_more_models}
\end{figure*}

In this section, we discuss the technical details in the computation of the DSFF, including the unfolding procedure and the angular dependence, and we demonstrate the GinUE  universality of the long-range spectral correlations of the dual spectra by computing DSFF for two-site Haar-random unitary gate (Fig.~\ref{fig:dsff_more_models} left), TI-BWM with $q=2$ (Fig.~\ref{fig:three_panels}a in the main text), TI-RPM with $q=3$ (Fig.~\ref{fig:dsff_more_models} middle), and TIF-BWM with $q=2$ (Fig.~\ref{fig:dsff_more_models} right).

\begin{figure*}[t]
\begin{minipage}[t]{0.31\textwidth}
\includegraphics[width=1.1\linewidth,keepaspectratio=true]{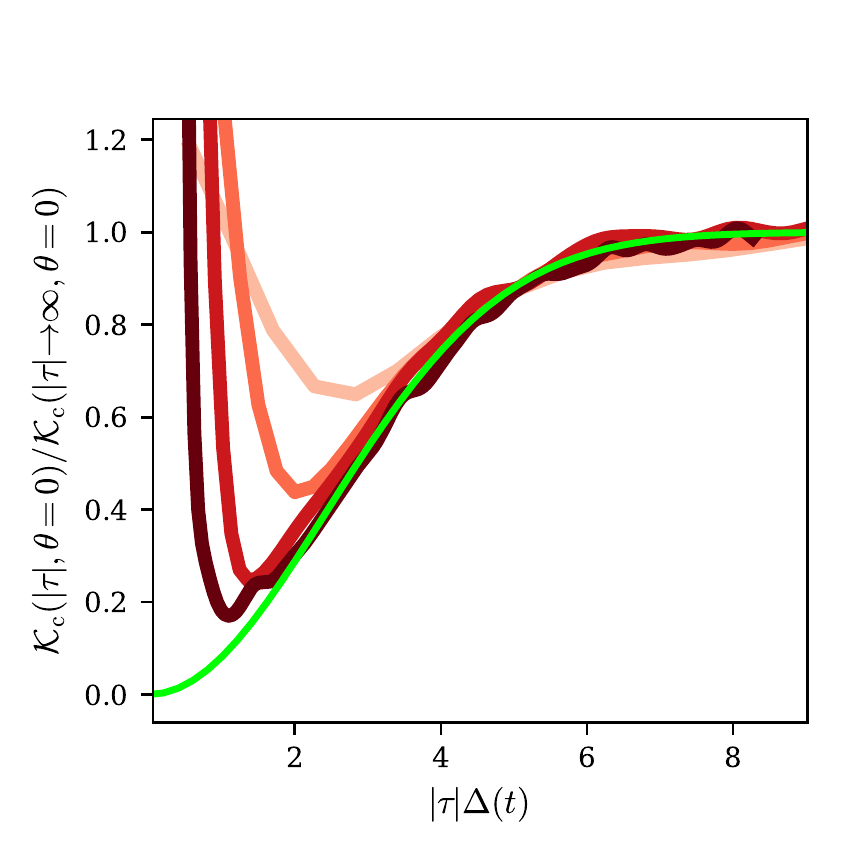}
\end{minipage}
\hspace*{\fill} 
\begin{minipage}[t]{0.31\textwidth}
\includegraphics[width=1.1\linewidth,keepaspectratio=true]{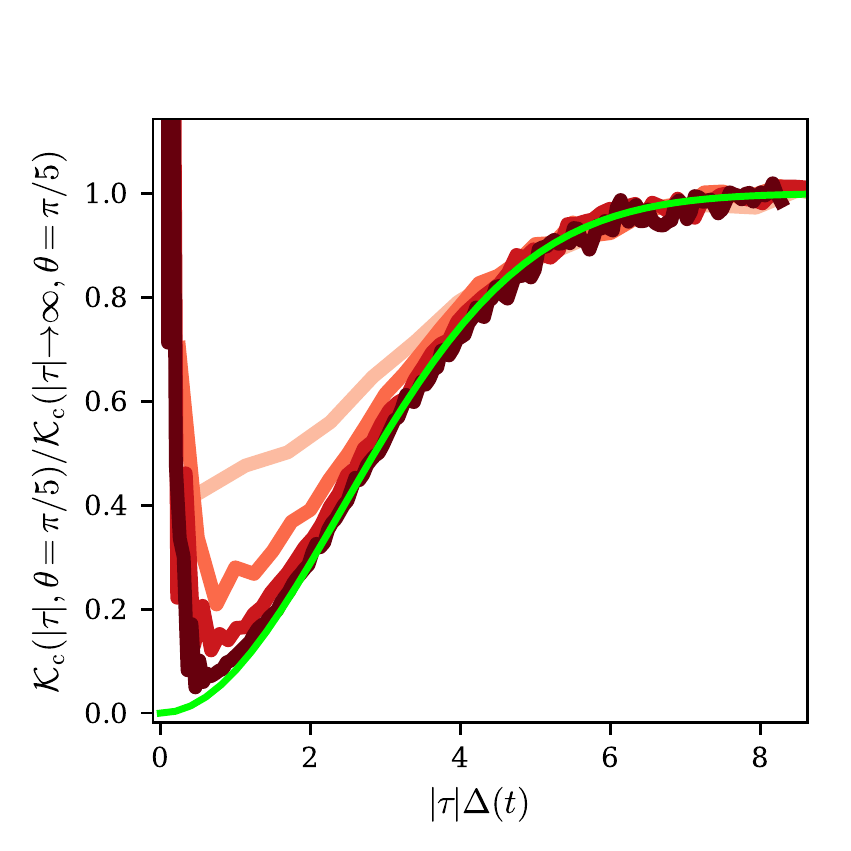}
\end{minipage}
\hspace*{\fill} 
\begin{minipage}[t]{0.31\textwidth}
\includegraphics[width=1.1\linewidth,keepaspectratio=true]{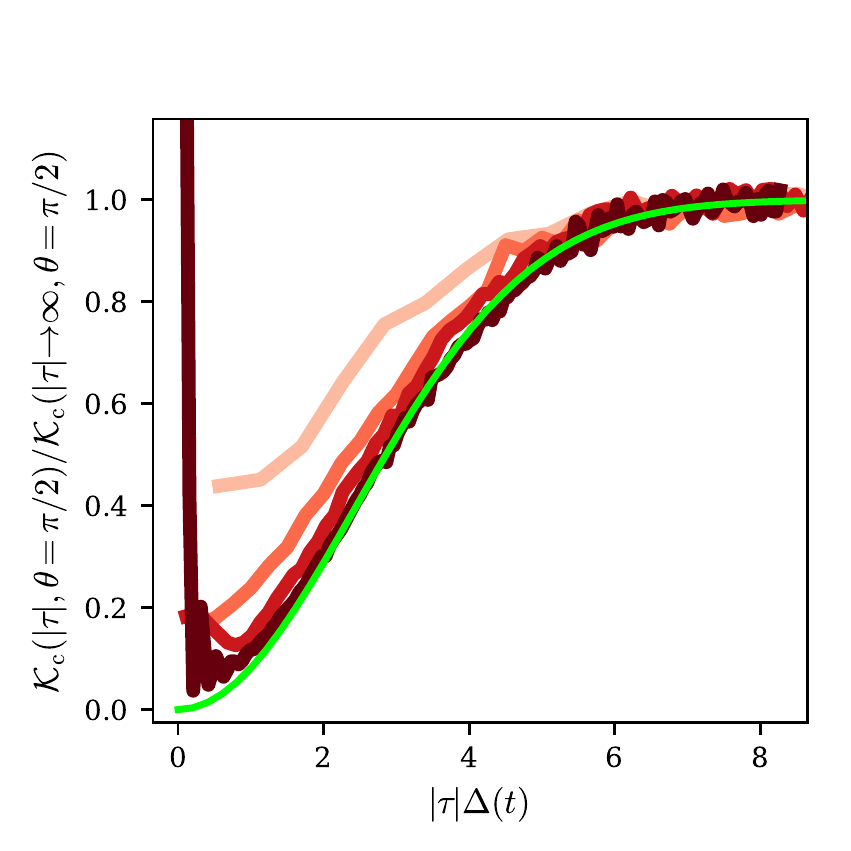}
\end{minipage}
\caption{$\theta$-dependence of connected DSFF $\dsffc(|\tau|, \theta)$ for some choice of angles for TI-BWM with $q=2$, $t=6,8,10,12$ (increasing in darkness), $\theta=0$ (left), $\theta=\pi/5$
(middle) and $\theta=\pi/2$ (right). The initial dip varies with the angle. However, the location of the dip drifts toward the origin in thermodynamic limit, and the DSFF approaches the GinUE DSFF behaviour.    } \label{fig:dsff_bwm_theta}
\end{figure*}

The quadratic ramp of (connected) DSFF is sensitive to the variation of density of states (DOS) across the complex plane --- different local region of eigenvalues contribute to a ramp-plateau DSFF curves that plateau at different effective Heisenberg times, and sum to a smeared-out non-RMT-like DSFF behaviour.
Therefore, the study of spectral correlation of complex spectra typically requires the complex generalization of unfolding~\cite{Mehta}, which is considerably more delicate than the real counterpart since the unfolding transformation needs to be conformal. 
Specifically, 
the (connected) DSFF, defined in Eq. \ref{eq:dsff_def_a}, is computed for dual circuits after performing the unfolding procedure described in Appendix \ref{unfolding}. 
%
%
%
%
%
%
%
For example for the DSFF of TI-BWM in Figure \ref{fig:three_panels}a, the unfolded eigenvalues are $\xi=z^{\frac{1}{t+1}}$.
The level spacing of original eigenvalues scale as $\Delta(t)\sim N^{-1/2}=q^{-t/2}$, and after unfolding scale as $\Delta(t) \sim 1/\sqrt{(t+1)N}$, inferred by looking at the collapse of DSFF with different dual system size $t$.
Averages are performed around  $10000$ to $15000$ realizations. Note that in Figure \ref{fig:three_panels}a and wherever not specified, the angle $\theta$ for which DSFF is computed, is chosen to be  $\theta=\pi/5$ away from the axes. 
The DSFF for different $\theta$ for TI-BWM are shown in Figure \ref{fig:dsff_bwm_theta}, with the same unfolding procedure used. 
Because of rotational symmetry of eigenvalues, we see similar approaches to $\dsffgin(|\tau|)$ for different $\theta$, except for the initial dips of the DSFF appear at different $\tau$. 
However, for all of them, the dip approaches zero for larger system sizes. That scale also depends on the cutoff chosen for the eigenvalues. 

%
Lastly, the universality of dual spectral statisticss obtained through DSFF are further argued for in Figure \ref{fig:dsff_more_models}, by looking at the  DSFF of the dual of two-site unitary gate with dimension $q^2\times q^2$ \ref{app:mod_2site}, arguably the simplest model with space-time duality. 
We see that the DSFF of the dual of this ensemble approaches the DSFF of GinUE as $q$ increases.
%
Note that the density of eigenvalues for $q>2$ is roughly flat (Fig.~\ref{fig:eigs_abs_dist}), and no unfolding is required. Consqeurently, there are no dips for  small $|\tau|$ in this case.

\section{SFF for GinUE}
\subsection{Exact evaluation of SFF for GinUE}
%
In this section, we evaluate the SFF for the GinUE. The definition of SFF is reproduced as follows.
\begin{equation}\label{eq:app_sff_def}
\KGinUE(L) 
=
\left\langle
\left| \Tr[\VG^L] \right|^2 
\right\rangle
\end{equation}
where $\VG$ is a $N$-by-$N$ matrix drawn from the GinUE with variance $\sigma^2 = v/N$, and $\langle \cdot \rangle$ is the ensemble average over the GinUE.
The joint probability distribution function of eigenvalues of GinUE is known exactly, and the correlation function of eigenvalues can be expressed in terms of the kernel \cite{ginibre_1965},
$
\kk(z_1, z_2) = 
\frac{N}{\pi \varr}
e^{-\frac{N}{2 \varr}(|z_1|^2 + |z_2|^2)}
\sum_{\ell= 0}^{N-1} \frac{1}{\ell !} \left( \frac{N z_1 z_2^*}{ \varr} \right)^\ell
\;.
$
The 1-point correlation function, i.e. the density of states,
is given by $\langle  \rho(z) \rangle = k(z,z)$, and the kernel is normalized such that $\int d^2 z \, \langle  \rho(z) \rangle = \int d^2 z \, k(z,z)= N$. Note that the DOS is isotropic, 
and is asymptotically, as $N\to\infty$, flat on a unit disc $|z|<1$ and vanishing outside.
The 2-point correlation function is
$
\langle 
\rho(z_1) \rho(z_2) 
\rangle 
=
k (z_1, z_1) \delta(z_1 - z_2) 
+ k (z_1,z_1)k (z_2,z_2)
-
\left| k (z_1,z_2)\right|^2
$.
We will refer to the above three terms as the contact, disconnected and connected term respectively.
The SFF of GinUE can now be written as 
\be
\KGinUE(L)=
\iint d^2 z_1 d^2 z_2
\;
\langle 
\rho(z_1) \rho(z_2) 
\rangle  
\;
z_1^L z_2^{*L} \;.
\ee
The contact term can be evaluated as follows
\be\label{eq:contact_term}
\int d^2 z \; |z|^{2L} k(z, z) = \sum_{\ell =0}^{N-1} \frac{v^L}{N^L \ell! } (L+\ell)! 
= \frac{ v^L (N+L)!}{ N^L (L+1) (N-1)!} \;,
\ee
where we have used the identity $\int_0^\infty d|z| \, |z|^{2p+1 } e^{-  N|z|^2 /v} = \frac{1}{2} \left( \frac{\varr}{N} \right)^{p+1} \Gamma(p+1)$. \eqref{eq:contact_term} is indeed $N$ at $L=0$ as expected. The disconnected term is given by 
\be\label{eq:disconnected_term}
 \left| \int d^2 z \; z^{L} k(z, z) \right|^2 = N^2 \,  \delta_{L,0}  \;.
\ee
Lastly, the connected term can be evaluated as 
\be\label{eq:connected_term}
\begin{aligned}
 -\int d^2 z_1 \,  d^2 z_2  \, z_1^L z_2^{*L} \, \left| k (z_1,z_2)\right|^2  
 =&
 -
 \int d^2 z_1 \,  d^2 z_2  \, 
 \frac{4N^2}{v^2}   
e^{-\frac{N}{2 \varr} (|z_1|^2 + |z_2|^2)}
\sum_{\ell= 0}^{N-L-1 } 
\frac{ 
N^{L+2\ell} \, |z_1|^{ 2(L+\ell)+1 } \, |z_2|^{2(L+\ell)+1  } 
}{
\varr^{L+2\ell}  \,
\ell! \, (L+ \ell)!
} 
\\
=& - \sum_{\ell=0}^{N-L-1} \frac{v^L \, [(L+\ell)!]^2
}{
N^L  
\ell! (L+\ell)!
}
= -
\frac{v^L \, N!}{N^L \, (L+1) (N-L-1)!}
   \;.
   \end{aligned}
\ee
%
Together, we arrive
\be
\begin{aligned} \label{eq:app_gin_sff}
\KGinUE(L) 
=  N^2 \delta_{L,0}+
\frac{ \varr^L \left((L+N)!-\frac{N! (N-1)!}{ (N-L-1)!}\right)}{N^L (L+1) (N-1)!} \;.
\end{aligned}
\ee
%
We reproduce the large $N$ expansion provided in the main text below as 
\be\label{eq:gin_largeN}
\KGinUE(L) = \varr^L
L
\left[
1+
\frac{(L-2) (L-1) L(L+1)}{ 24 N^2}    
+O\left( \frac{L^8}{N^{4}} \right)
\right]
\ee
which is verified by the diagrammatical approach in large-$N$  below.

\subsection{Diagrammatical approach}
\begin{figure}[H]
\centering
\includegraphics[width=0.6\textwidth]{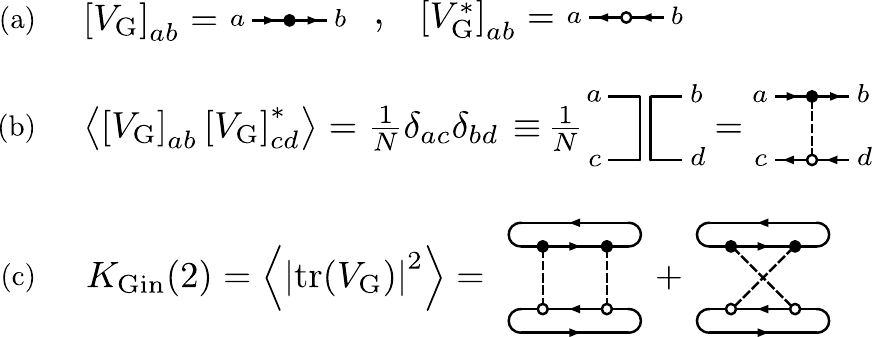}
\caption{Diagrammatical representation of (a) the matrix element of matrix $\VG$ drawn from the Ginibre ensemble, (b) a constraction, and (c) the (ensemble-averaged) SFF $\KGinUE (2)$.
\label{Fig:ginue_diag}}
\end{figure}

\begin{figure}[H]
\centering
\includegraphics[width=0.95\textwidth]{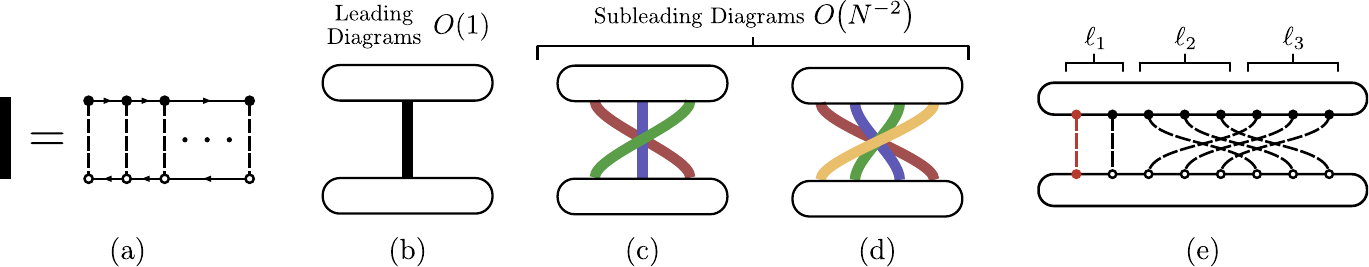}
\caption{(a) The thick line notation for a group of parallel contractions. (b) Leading diagrams in $N$ of $\KGinUE$. (c, d) Subleading diagrams of $\KGinUE$ of order $O\left(N^{-2} \right)$. Thick lines are colored to guide the eyes.
\label{Fig:ginue_sff_diag}}
\end{figure}

%
%
%
%
%
The diagrammatical approach of GinUE is useful as analytical tool for variations of GinUE-like models, and also 
as a consistency check for the exact computation of SFF \eqref{eq:app_gin_sff} and \eqref{eq:gin_largeN} in large $N$.
The diagrammatical approach is constructed by representing a matrix $\VG$ drawn from the GinUE as in Fig.~\ref{Fig:ginue_diag}a, and a contraction as in  in Fig.~\ref{Fig:ginue_diag}b. 
The ensemble average of the SFF  of the GinUE at $L=2$ is given as  an example in Fig.~\ref{Fig:ginue_diag}c. 
%
%

Now we verify the result in \eqref{eq:app_gin_sff} and \eqref{eq:gin_largeN} using the diagrammatical approach in large $N$. Note that we fix $v=1$ in this section. 
In large $N$, ensemble-averaged observables are dominated by diagrams with the most number of solid-line loops in the notation given in  Fig.~\ref{Fig:ginue_diag}b (see also ~\cite{cdc1}), since each loop corresponds to a factor of $N$. 
The leading diagrams are the well-known ladder diagrams given in \ref{Fig:ginue_sff_diag}b,
and it is shown in \cite{ceren} that the subleading diagrams of SFF are of the form \ref{Fig:ginue_sff_diag}c and d. 
Firstly, We can count the degeneracy of the ladder diagran following~\cite{cdc1}. Suppose there are $L$ nodes in the upper trace and $L$ nodes in the lower trace.
The top left node can be chosen to contract to any one of the bottom nodes, and the rest of the nodes on the top loop have to be contracted such that a ladder is formed. Hence there are $L$ choices, i.e. the degeneracy $D_{\mathrm{a}}$ of diagram \ref{Fig:ginue_sff_diag}a is 
\be
D_{\mathrm{b}}= L
\ee
Secondly, to count diagrams in  \ref{Fig:ginue_sff_diag}c, consider a realization of \ref{Fig:ginue_sff_diag}c given in Fig.~\ref{Fig:ginue_sff_diag}e.
We use the labels $\ell_1$, $\ell_2$ and $\ell_3$ to denote the number of contractions in the three groups of parallel contractions. 
We first consider all possible diagrams with the top left node connecting with the bottom left node, colored in red in Fig.~\ref{Fig:ginue_sff_diag}e. 
$\ell_1$ can take values from $1$ to $L-2$.
$\ell_2$ can take $L-\ell_1 -1$ choices.
For each choices of $\ell_2$, $\ell_3$ is fixed since $\sum_{i=1}^3 \ell_i = L$.
Furthermore, we have to account for the fact that the red contractions can be one of any contractions in the group of $\ell_1$ contractions.
Finally, we unfix the contraction of the top left node and allow it to contract to any one of the $L$ nodes in the bottom loop. This gives an additional factor of $L$. 
Therefore, we have degeneracy $D_{\mathrm{c}}$ given by
\be
D_{\mathrm{c}}=L \sum_{\ell_1= 1}^{L-2} \ell_1 (L - \ell_1 - 1)  = \frac{(L-2) (L-1) L^2}{6}
\; .
\ee
Thirdly, we can repeat the same counting argument for \ref{Fig:ginue_sff_diag}d, and obtain
\be
D_{\mathrm{d}}=L \sum_{\ell_1= 1}^{L-3} \ell_1  \sum_{\ell_2=1}^{L-\ell_1 - 2} (L - \ell_1 - \ell_2 - 1)  = \frac{(L-3)(L-2) (L-1)L^2}{24} 
\; .
\ee
Together, up to $O(N^{-2})$, we have
\be
\KGinUE(L) = D_{\mathrm{b}} + \frac{1}{N^2}( D_{\mathrm{c}} + D_{\mathrm{d}}) +O\left(\frac{1}{N^4}\right)
=
 L+\frac{(L-2) (L-1) L^2(L+1)}{24 N^2}+O\left(\frac{1}{N^4}\right) \;,
\ee
which verifies Eq.~\eqref{eq:gin_largeN}.


\section{SFF for Floquet GinUE model}\label{app:sff_floq}
In this appendix, we compute the SFF of a Ginibre model for the DTM of Floquet but spatially random MBQC systems, and thereby demonstrate that the emergence of GinUE-like behaviour persists beyond TI systems.
The GinUE-like model is given by $\mathcal{V}(t,L) = \prod_{r=1}^L \VG(t,r)$, such that $\VG(t,r) = T \VG(t,r) T^{-1} $, where $T$ translates the dual system over one period, and where we impose spatial randomness, i.e. $\VG(r) \neq \VG(r')$ for $r\neq r'$. 
For simplicity, we assume invariance under one site translation, with
$T \ket{\mathbf{s} = s_1s_2\ldots s_t} = \ket{ s_2s_3\ldots s_ts_1}$.  Generalization to longer unit cells can be straightforwardly obtained.
%
As described in the main text, we restrict the Hilbert space to the set of  computational basis $\{ \ket{\mathbf{s}}\}$ translational invariant with only period $t$.
We note that the fraction of configurations with maximal period goes to $1$ for large $t$ (and/or obviously for large $q$). 
We demand $\VG$-s to have Ginibre-like correlation given by 
\be \label{eq:gin_corr2}
\left\langle [\VG]_{\bs \bs'} [\VG]^
*_{\bp \bp'} \right\rangle 
=
\frac{1}{\Nb}
\sum_{\tau, \tau'}
\delta_{ \mathbf{s} T^\tau(\mathbf{p}) } \delta_{\mathbf{s}' T^{\tau'}(\mathbf{p}') }
J(\tau - \tau')\;,
\ee 
where $\Nb$ denotes the number of orbits, and 
$J(\tau)$ determines the correlation between matrix elements.
%
%
Recall that $s$ is a computational basis string, translational invariant with period $t$.
It is convenient to represent each basis state $s = (\tilde{\mathbf{s}}, \tau)$ in terms of a representative  $\tilde{\mathbf{s}}$ of the orbit and a translation parameter $\tau$ such that $s= T^{\tau} \tilde{s}$.
In this representation, the SFF of this Floquet Ginibre model  can be evaluated as the ladder diagram given in \ref{Fig:ginue_sff_diag}a and b. 
%
(Due to spatial randomness, there is only a single ladder diagram since $\VG(r)$ can only correlate with $\VG^*(r)$ for a given $r$.)
By summing over the $s$ degrees of freedom, the SFF can be written as 
\be \label{eq:k_floq_ft}
K(t,L) =  \left\langle 
\left| \Tr[\VG(t,L)] \right|^2 
\right\rangle =
\sum_{\{ \tau \}}
\prod_{i =1}^L J(\tau_{i} - \tau_{i+1}) = 
  \sum_\omega [\hJ(\omega)]^L \; . 
\ee
where we have used the Fourier transform
\be
\hJ(\omega ) = \sum_{\tau} J(\tau )e^{-\frac{i 2\pi  \omega \tau}{t}}
\;, \qquad \qquad 
J(\tau) = \frac{1}{2 \pi}\sum_{\omega} \hJ(\omega )e^{\frac{ i 2\pi \omega \tau}{t}} \;.
\ee
Now we consider a general class of correlation with  $J(\tau)$ of form  
\be\label{eq:J_form}
J(\tau) = \delta_{\tau ,0} + f(t) h(\tau)
\ee
where $f(t)$ is some function that decays from 1 to 0 in $t$. 
Together with the first term in \eqref{eq:J_form}, the second term ensures that there are $t$ ``domain wall'' contribution, which is expected to dominate the SFF in late time~\cite{cdc1, cdc2}. 
For simplicity, we further assume that the second term has a factorized form in $t$ and $\tau$. 
%
Lastly, note that in the convention of \eqref{eq:J_form}, $h(\tau =0) = 0$.
The Fourier transform of \eqref{eq:J_form} is given by
\be\label{eq:J_form_ft}
\hJ(\omega) = 1  + f(t) \hh \left( \frac{2\pi \omega }{t} \right) \equiv  1  + f(t) \hh \left( \tomega \right)  \; . 
\ee
We will explore two cases of $J(\tau)$, namely, the fast-decaying or ``short-range'' and slow-decaying or ``long-range'' $J(\tau)$ in $\tau$, and derive $K(L)$ and its scaling forms.
In particular, when $J(\tau)$ is long-range, we recover the scaling form and an emergent Potts model, recovering the results in \cite{chan2021manybody, cdc2}, and thereby showing the emergence of Ginibre-like ensemble beyond TI MBQC systems.


\subsection{Long-range $J(\tau)$}
Suppose we take $J(\tau)$ to be long range, i.e. slow-decaying in $\tau$. 
For sufficiently large $t$ where $f(t)$ small, we substitute \eqref{eq:J_form_ft}  into \eqref{eq:k_floq_ft} and write 
\be
\kfgin(t, L) = \sum_\omega [1+ f(t)\hh(\tomega) ]^L 
\approx 
e^{f(t) \hh(0) L} +
\sum_{\tomega \neq 0} e^{f(t) \hh(\tomega) L}
\approx 
e^{f(t) \hh(0) L} +
\frac{t-1}{2\pi} \int d\tomega \,   e^{f(t) \hh(\tomega) L}\;.
\ee
Taking $x = f(t) \hh(0) L \equiv L/ \Ls(t)$ with $\Ls \equiv [f(t) \hh(0)]^{-1} $, we have
\be\label{eq:k_longrange_step}
\kfgin(t, L) 
\approx 
e^{x} +
\frac{t-1}{2\pi} \int d\tomega \,   e^{
x  \hh(\tomega)  / \hh(0) }
\approx
e^{x} +
t -1 + \frac{(t-1) x}{2\pi}
 \int d\tomega \, 
 \frac{ \hh(\tomega)  }{ \hh(0) }   \; .
\ee
To evaluate the last term, we note that 
\be
0 = h(\tau= 0) =  \frac{t}{2\pi} \int d\tomega \, h(\tomega) = \hh(0) +    \frac{t}{2\pi} \int d\tomega \, h(\tomega) \;.
\ee
Substituting this into \eqref{eq:k_longrange_step}, we have SFF
\be
\kfgin(t, L) 
\approx
e^{x} +
t -1 -x   \; .
\ee
and the scaling form
\be
\kappafgin(t, L)  
= 
\lim_{\substack{L,t \to \infty\\ x =L/\Ls(t) }}
   \kfgin
   -  t
   =
  e^{x} -x  -1
  \; .
\ee
Consequently, we recovered the results derived in \cite{chan2021manybody}, i.e.
\be 
  \kappaginueB(x) 
  =\kappa_{\mathrm{F}-\mathrm{MBQC}} (x)  
\;.
\ee

\subsection{Long-range $J(\tau)$ example: Emergent Potts model}
As an example of long-range $J(\tau)$, we take $J(\tau)$ to be
\be\label{eq:J_potts}
J(\tau) = \delta_{\tau,0} + f(t) (1- \delta_{\tau,0}) \;,
\ee
with
$f(t)$ vanishing for $t > \ts$, the analogue of {Thouless time} for this Floquet-GinUE model, 
we recover exactly the SFF of MBQC system evaluated at infinite-$q$ in \cite{cdc2}.
The fourier transform of \eqref{eq:J_potts} gives
\be
\hJ(\omega) = (1-f) + tf \delta_{\omega,0}  \;,
\ee
and therefore, 
\be
K_{\mathrm{F}-\mathrm{Gin}} =[1+(t-1)f]^L + (t-1)(1-f)^L \;.
\ee
This coincides with the SFF result for Floquet MBQC systems in \cite{cdc2}, if $f(t) = e^{-\epsilon t}$ for some parameter $\epsilon$.
Taking the scaling limit where $L$ and $t$ are sent large, and where $x = L t f(t) $ fixed, we have 
\be \label{eq:conj}
  \kappaginueB(x)= 
 \lim_{\substack{L,t \to \infty\\ x =L t f(t)}}  K_{\mathrm{F}-\mathrm{Gin}} -t  = e^{x} - x -1
\;,
\ee
which again coincides with the SFF scaling form for Floquet MBQC systems in \cite{cdc2}.
In summary, we have
\be
K_{\mathrm{F-Gin}}(t,L) = Z_{ \mathrm{Potts}} 
= K_{\mathrm{F-MBQC}}(t,L)
\;,
\ee
where $Z$ is the partition function of a classical $t$-state ferromagnetic Potts model with  Boltzmann weight 
$\mathcal{B}(\sigma_i, \sigma_{i+1})= \delta_{\sigma_i, \sigma_{i+1}}
+
f(t)(1-\delta_{\sigma_i, \sigma_{i+1}}) 
$ and $\sigma_i$ labels the Pott state at site $i$. 
%
Consequently, taking the scaling limit, we have the exact identity
\be 
  \kappaginueB(x) 
  =\kappa_{\mathrm{F}-\mathrm{MBQC}} (x)  
\;.
\ee
%
%

\subsection{Short-range $J(\tau)$}
Suppose we take $J(\tau)$  to be short range, i.e. fast-decaying in $\tau$. For sufficiently large $t$ where $f(t)$ small, we substitute \eqref{eq:J_form_ft}  into \eqref{eq:k_floq_ft} and write 
\be
\kfgin(t, L) = \sum_\omega [1+ f(t)\hh(\tomega) ]^L \approx \sum_\omega e^{f(t) \hh(\tomega) L}
\approx \frac{t}{2\pi} \int d\tomega \,   e^{f(t) \hh(\tomega) L}\;.
\ee
Casting the above equation in the scaling form, we write 
\be
\begin{aligned} 
   \kappa_{\mathrm{F}-\mathrm{Gin}} (x)=   \lim_{\substack{L,t \to \infty\\ x =L/\Ls(t) }}
   \frac{
   \kfgin
     }{
     t
     }
      \;, 
\end{aligned}
\ee
where $\Ls(t) \equiv 1/f(t)$. Importantly, this scaling form does not coincide with the analogous scaling form derived for long-range $J(\tau)$, and the results in \cite{cdc2,chan2021manybody}.

%
%
\subsection{Short-range $J(\tau)$ example with nearest-neighbour coupling}
As an example of short-range $J(\tau)$, we take  
\be\label{eq:J_short}
J(\tau) = \delta_{\tau,0} + f(t) (\delta_{\tau,1}+  \delta_{\tau,-1}) \;,
\ee
with Fourier transform given by 
\be
\hJ(\omega) = 1 + 2f \cos\left( \frac{2\pi \omega }{t} \right) \;,
\ee
The SFF can be computed as 
\be
\kfgin(t, L) = \sum_\omega \left[1+ 
2f \cos\left( \tomega \right)
\right]^L 
\approx \frac{t}{2\pi} \int d\tomega \,   e^{2 f(t) \cos(\tomega) L}
= t I(2f(t) L)
\;,
\ee
where $I$ is the Bessel function.



\section{SFF for TI Floquet GinUE model}
In this appendix, we compute the SFF of a Ginibre model for the DTM of TIF MBQC systems.
The TIF Ginibre model is given by $\mathcal{V}(t,L) = \prod_{r=1}^L \VG(t,r)$, with $\VG(t,r)$ satisfying $\VG(t,r) = T \VG(t,r) T^{-1} $, $\VG(t,r) = \VG(t,r')$ for all $r, r'$ and \eqref{eq:gin_corr}.
Recall that $T$ translates the dual system over one period.
We will use the same representation for the basis state $\mathbf{s} = (\tilde{\mathbf{s}}, \tau)$, as in Appendix \ref{app:sff_floq}, where $\tilde{\mathbf{s}}$  is a representative   of the orbit of $\mathbf{s}$, and  $\tau$ is a parameter such that $\mathbf{s}= T^{\tau} \tilde{\mathbf{s}}$.
In this representation, the matrix elements of $V$ is given by,
\be
V_{\mathbf{s} \mathbf{s}'}^{\tau \tau'} \equiv \bra{\mathbf{s}', \tau' }V \ket{\mathbf{s}, \tau} = V_{\mathbf{s} \mathbf{s}'}^{\tau-\tau'}
\ee
where the last equality is due to $\VG(t,r) = T \VG(t,r) T^{-1}$, and we have left the subscript G implicit. Using the Fourier transformation $\ket{\mathbf{s}, \tau} = (2\pi)^{-1/2} \sum_{\omega} e^{\imath 2\pi \tau \omega / t} \ket{\mathbf{s}, \omega}$ and
$\ket{\mathbf{s}, \omega} = (2\pi)^{-1/2} \sum_{t} e^{-\imath 2\pi \tau \omega / t} \ket{\mathbf{s}, t}$, we write
\be
\bra{\mathbf{s}', \omega' }V \ket{\mathbf{s}, \omega} 
= 
\delta_{\omega, \omega'}
\sum_{\tau} 
e^{-\imath 2\pi \tau \omega / t}
V_{\mathbf{s} \mathbf{s}'}^{\tau}
\ee
We can then rewrite the covariance,
\be
\left\langle
V^{\tau}_{\mathbf{s}_1 \mathbf{s}_2} V^{* \tau'}_{\mathbf{s}_3 \mathbf{s}_4}
\right\rangle
= 
\frac{J(\tau - \tau')}{\Nb} \delta_{\mathbf{s}_1,\mathbf{s}_2} \delta_{\mathbf{s}_3,\mathbf{s}_4}
\ee
in the frequency space as 
\be\label{eq:app_freq_space}
\left\langle
V^{\omega}_{\mathbf{s}_1 \mathbf{s}_2} V^{* \omega'}_{\mathbf{s}_3 \mathbf{s}_4}
\right\rangle
=
\delta_{\omega \omega'} 
\frac{\hJ(\omega)}{\Nb} \delta_{\mathbf{s}_1,\mathbf{s}_2} \delta_{\mathbf{s}_3,\mathbf{s}_4}
\ee
where $
V_{\mathbf{s} \mathbf{s}'}^{\omega} \equiv 
\bra{\mathbf{s}', \omega }V \ket{\mathbf{s}, \omega} $.
Eq.~\eqref{eq:app_freq_space} means that different frequency sectors are statistically decoupled. 
Noting that matrix elements of $V$ in \eqref{eq:app_freq_space} behave like matrix elements of a Ginibre matrix, with variance $v/N \to \hat J(\omega)/\Nb$.
We can thus evaluate $\ktifgin$ for this model, using Eq.~\eqref{eq:kginue} within each frequency sector and  obtain for $L \neq 0 $,
\begin{equation}
\begin{aligned}
\ktifgin(t,L) 
& = \frac{ \left((L+\Nb)!-\frac{\Nb! (\Nb-1)!}{ (\Nb-L-1)!}\right)}{\Nb^L (L+1) (\Nb-1)!} 
\sum_\omega \left[ \hat J(\omega) \right]^L 
\\
&= 
\KGinUE (\Nb, L) \kfgin(t,L)  
\\
&
\sim L \, \kappaginueC \left(\frac{L}{\Lsc}\right) \left[\kappaginueB \left(\frac{L}{\Lsb}\right) + t \right] \;,
\end{aligned}
\end{equation}
where in the last line, we have used the results in Eqs.~(\ref{eq:kc_scaling_dgt1},\ref{eq:kappafgin}).

\section{SFF: Numerical results}
\begin{figure}[H]
\centering
\includegraphics[width=0.5\textwidth]{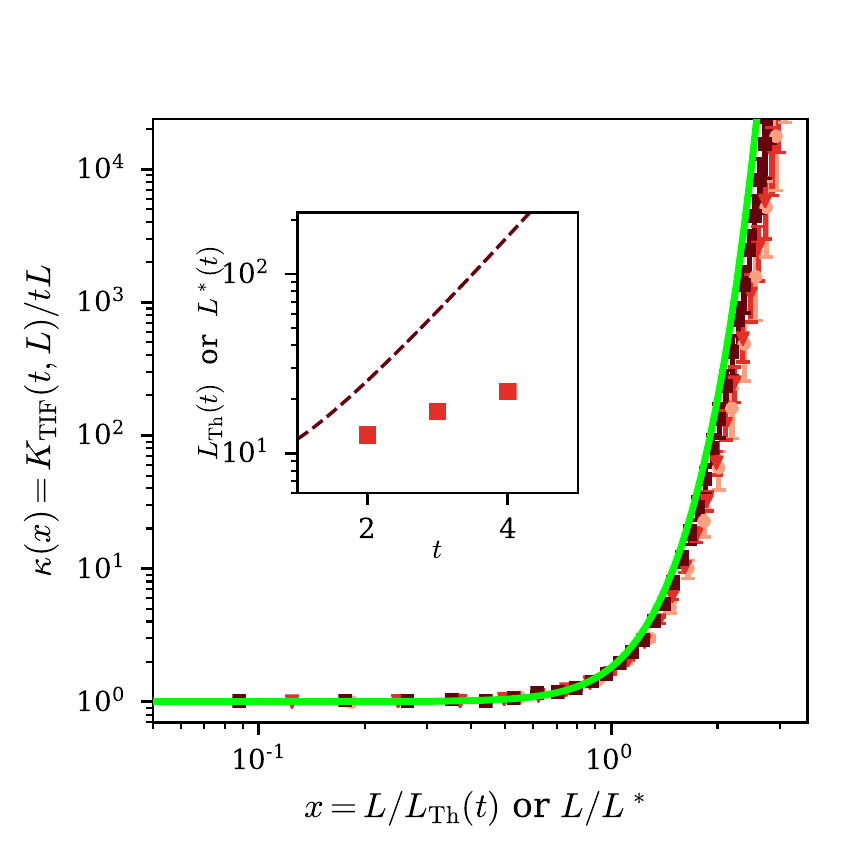}
\caption{
Scaling collapse of $\kd (t,L)/tL$ against $x$ for TIF-BWM with $q=3$, $t=2,3,4$ (reds), also shown in Fig.~\ref{fig:three_panels}e.
The scaling form $\kd/tL$ agrees with $\kappa_{\mathrm{GinUE}}(x)$ in Eq. \ref{eq:caseD_collapse1}. The inset shows the corresponding $\Lth(t)$ and $L^*(t)$  against $t$. 
}
\label{fig:caseD_eq13}
\end{figure}

In this section, 
we provide further scaling collapses of the TIF-BWM  as a representative model of TIF MBQC systems.
From the numerical fitting in \ref{fig:three_panels}f, we have $\Lsb < \Lsc$ for TIF-BWM. This motivates us to plot the scaling collapse according to Eq.~\eqref{eq:caseD_collapse1} in Fig.~\ref{fig:caseD_eq13}, and we see that agreement is obtained with the scaling function provided in Eq.~\eqref{eq:caseD_collapse1}.
%
%
%
Lastly, note that $\Lth(t)$ is computed by looking at the intersection points of $K_{\mathrm{TI}}(t,L)/L$ vs $L$ with a horizontal line close to $1$ (see also the Appendices in \cite{chan2021manybody}), and averaged over around $10000$ to $15000$ realizations.

\end{document}